\documentclass[useAMS,usenatbib,fleqn]{mnras}
\usepackage{ae,aecompl}
\usepackage[dvipsnames]{xcolor}

\title[Observational Bias and YMC Characterisation I.]
{Observational Bias and Young Massive Cluster Characterisation \\I. 2D Perspective Effects}
\author[Buckner et al.]
{Anne S.M. Buckner\thanks{E-mail:
a.buckner@exeter.ac.uk}$^{1}$, Kong You Liow$^{1}$, Clare L. Dobbs$^{1}$, Tim Naylor$^{1}$, Steven Rieder$^{2,1}$ \\
$^1$ School of Physics and Astronomy, University of Exeter, Stocker Road, Exeter, EX4 4QL, UK \\
$^2$ Geneva Observatory, University of Geneva, Chemin Pegasi 51, 1290 Sauverny, Switzerland \\}

\usepackage{amssymb}
\usepackage{amsmath}
\usepackage[pdftex]{graphicx}
\usepackage{epsfig}
\usepackage{multirow} 
\def\aap{{ A\&A}}

\def\apj{{ApJ}}
\def\apjl{{ApJL}}

\def\mnras{{MNRAS}}
\def\pasj{{PASJ}}

\def\apjs{{ApJS}}
\def\araa{{ARA\&A}}
\def\apss{{Ap\&SS}}

\usepackage{ulem}

\begin{document}
\label{firstpage}
\date{Accepted 2022 May 5. Received 2022 May 5; in original form 2022 February 8}

\pagerange{\pageref{firstpage}--\pageref{lastpage}} \pubyear{2012}

\maketitle

\begin{abstract}
Understanding the formation and evolution of high mass star clusters requires comparisons between theoretical and observational data to be made. Unfortunately, while the full phase space of simulated regions is available, often only partial 2D spatial and kinematic data is available for observed regions. This raises the question as to whether cluster parameters determined from 2D data alone are reliable and representative of clusters real parameters and the impact of line-of-sight orientation. In this paper we derive parameters for a simulated cluster formed from a cloud-cloud collision with the full 6D phase space, and compare them with those derived from three different 2D line-of-sight orientations for the cluster. We show the same qualitative conclusions can be reached when viewing  clusters  in  2D  versus  3D,  but  that  drawing quantitative conclusions when viewing in 2D is likely to be inaccurate. The greatest  divergence occurs in the perceived kinematics of the cluster, which in some orientations appears to be expanding when the cluster is actually contracting. Increases in the cluster density compounds pre-existing perspective issues, reducing the relative accuracy  and  consistency  of  properties  derived  from  different orientations. This is particularly problematic for determination of the number, and membership, of subclusters present in the cluster. We find the fraction of subclusters correctly identified in 2D decreases as the cluster evolves, reaching less than $3.4\%$ at the evolutionary end point for our cluster. 
\end{abstract} 


%~~~~~~~~~~~~~~~~~~~~~~~~~~~~~~~~~~~~~~~~~~~~~~~~~~~~~~

\section{Introduction}

Young massive clusters (YMCs) play a fundamental role in the evolution of galaxies. Active sites of large-scale star formation, they act as laboratories in which to study how stars form, as well as impact and spatially interact within their natal environment. They are particularly useful in the study of massive stars which typically do not form in isolation. (\citealt{2005A&A...437..247D}, \citealt{2007MNRAS.380.1271P}, \citealt{2012MNRAS.424.3037G}, \citealt{2017ApJ...834...94S}, \citealt{2018ARA&A..56...41M}, \citealt{2020MNRAS.495.1209R}). %form almost exclusively in clusters

Long the focus of intense observational study, intrinsic nebulosity associated with YMCs makes observational study difficult and datasets are almost always incomplete. 
 While the full extent of the effect this incompleteness has on the results of analyses performed using these datasets is not known, previous studies have indicated it is not trivial (\citealt{2009Ap&SS.324..113A}, \citealt{2012A&A...545A.122P}, \citealt{2022A&A...659A..72B}). A core issue is whether the parameters determined are representative of physical processes in the cluster or are attributable to observational biases and effects. For example, if only 2D spatial data is available is identified substructure real or a perspective effect? If a cluster is determined to be mass segregated, is this real or does it just appear to be due to perspective effects and/or heavy incompleteness in the lower mass population? The issue of structure identification is exemplified by studies such as \citet{Cantat-Gaudin2020}, who showed that some formerly identified Milky Way clusters are actually likely to be asterisms. \citet{Piecka2021} also argue that the 3D structure of clusters cannot be properly determined for clusters at distances $>500$ pc.

Another issue with observational datasets is that they can only give a ‘snapshot’ of a YMC at a single point in its evolution. Past and future spatial behaviours have to be extrapolated. But combined with the incompleteness problem, this is obviously difficult to achieve and the accuracy of any predictions questionable. For example, if no/few radial velocities are available, how trustworthy is a result that the cluster is expanding or contracting from proper motions alone? Observational studies serve a vital role in our understanding YMCs and informing star formation models, so it is critical that these questions are addressed so that their reliability can be ascertained. 

Simulated clusters provide a means to explore the properties of stellar clusters as examined with full 6 dimensional analysis. N-body simulations allow the evolution of clusters to be followed over time, and the actual evolution compared to that inferred from 2D or 3D velocity measurements at a single snapshot. At least at the formation stage, clusters will be embedded or at least associated with a natal gas cloud.  Simulations which include hydrodynamics can probe the effects of extinction, since the gas dynamics during the formation and evolution of clusters can be followed as well as the evolution of the stars. 

Previous works which have investigated properties of simulated clusters from an observational perspective include \citet{2019MNRAS.485.3124K}. They developed the MYOSOTIS tool\footnote{\url{https://github.com/zkhorrami/MYOSOTIS}}  which estimates a flux for each star, based on its mass, ages and metallicity, distance from the observer and properties of an assumed telescope. The code can also incorporate effects of extinction from gas surrounding the cluster, or a gas cloud between the cluster and the observer. \citet{2021arXiv211111805B} investigate how well properties of clusters such as the degree of clustering and mass segregation (using the INDICATE tool, \citealt{2019A&A...622A.184B}) would be measured using GAIA for a synthetic cluster placed inside a cloud at different distances, and with different binary fractions. Other work focuses on comparing the ages estimated from synthetic SEDs trace to the actual ages of stars in simulations \citep{2016A&A...587A..60F}. Studying the dynamics of clusters, \citet{2021A&A...647A.137J} compute the long term evolution of a Hyades-like cluster, investigating the origin of the shape of the tidal tails of the Hyades cluster. On extragalactic scales, \citet{2021arXiv211114875L} determine the 24 $\mu$m and H$\alpha$ emission from star clusters in a dwarf galaxy simulation, placing the galaxy at different distances to see how the estimated star formation rate varies.

In this series of papers we attempt to 
answer how well observations capture the properties and physics of clusters given limitations of 2D measurements and extinction. 
Here we explore the impact of 2D perspective effects on the derived properties of clusters. Using a simulation by \citet{liow_collision_2020} of a YMC formed via a cloud-cloud collision between two molecular clouds over a $\sim$0.96\,Myr timescale we follow the spatial and kinematic evolution of the cluster, first utilising the full 6D astrometric parameters, then with only 2D spatial/kinematic data along three different Lines-of-Sight (LoS). 

This paper is structured as follows. Section 2 details our YMC simulation set-up and cluster dataset. In section 3 we describe our analysis methods, and results are discussed in section 4. Our conclusions are presented in section 5.

%#########################################################################

\section{Data}
%~~~~~~~~~~~~~~~~~~~~~~~

\subsection{Simulation setup}
For the main part of this paper, we use a simulation based on the standard speed, low turbulence and high density cloud-cloud collision model in \cite{liow_collision_2020}. This model is chosen because it produces a compact and roughly spherical cluster at the collision site, which is visually similar to realistic clusters, yet the collision velocity is high enough to strongly determine the location of the stars, and induces highly asymmetric velocities during cluster formation. The mass of the main central cluster formed in the simulation is $\approx 1.8 \times 10^4$ M$_{\odot}$, similar to a YMC. In \citet{2020MNRAS.496L...1D} we argue that cloud-cloud collisions may be one way of producing massive clusters in short timescales, and are in any case more representative of conditions in more extreme environments in our or other galaxies. To check our results are representative, and not dependent on this one model we have chosen, we also perform analysis on the low speed, medium density cloud-cloud collision model, labelled as `L20Turbulent' in \cite{liow_grouped_2021}, which we describe in Appendix\,\ref{results_c1020}. This model is a rerun of the model of the same initial conditions in \cite{liow_collision_2020}, i.e. the model that created Cluster L4, but individual stars are evolved instead of just sink particles. 

In this previous paper we ran this simulation using \texttt{PHANTOM} \citep{price_phantom_2018}. However since this publication, we have developed the \texttt{Ekster} code \citep{rieder_ekster_2021}. The \texttt{Ekster} code converts sink particles to star particles, sampling from an IMF, which in contrast to most numerical simulations allows us to model the full stellar population of the cluster. Hence to carry out the work we present here, we reran the simulation from \cite{liow_collision_2020}, which used sink particles, but instead used \texttt{Ekster} in order to obtain individual stars. The \texttt{Ekster} code combines multiple physical processes including gas hydrodynamics, gravitational dynamics and stellar evolution via the \texttt{AMUSE} interface \citep{amuse_2018}. \texttt{PHANTOM} \citep{price_phantom_2018}, a smoothed particle hydrodynamics (SPH) code for astrophysics, is used to simulate the gas hydrodynamics in \texttt{Ekster}. 
\texttt{PeTar} \citep{wang_petar_2020} is used to calculate the gravitational interaction between the stars.
Lastly, \texttt{Ekster} uses the parametric stellar evolution code \texttt{SeBa} \citep{portegies_zwart_seba1_1996}.

\begin{figure*}
\centering
   \includegraphics[width=0.8\textwidth]{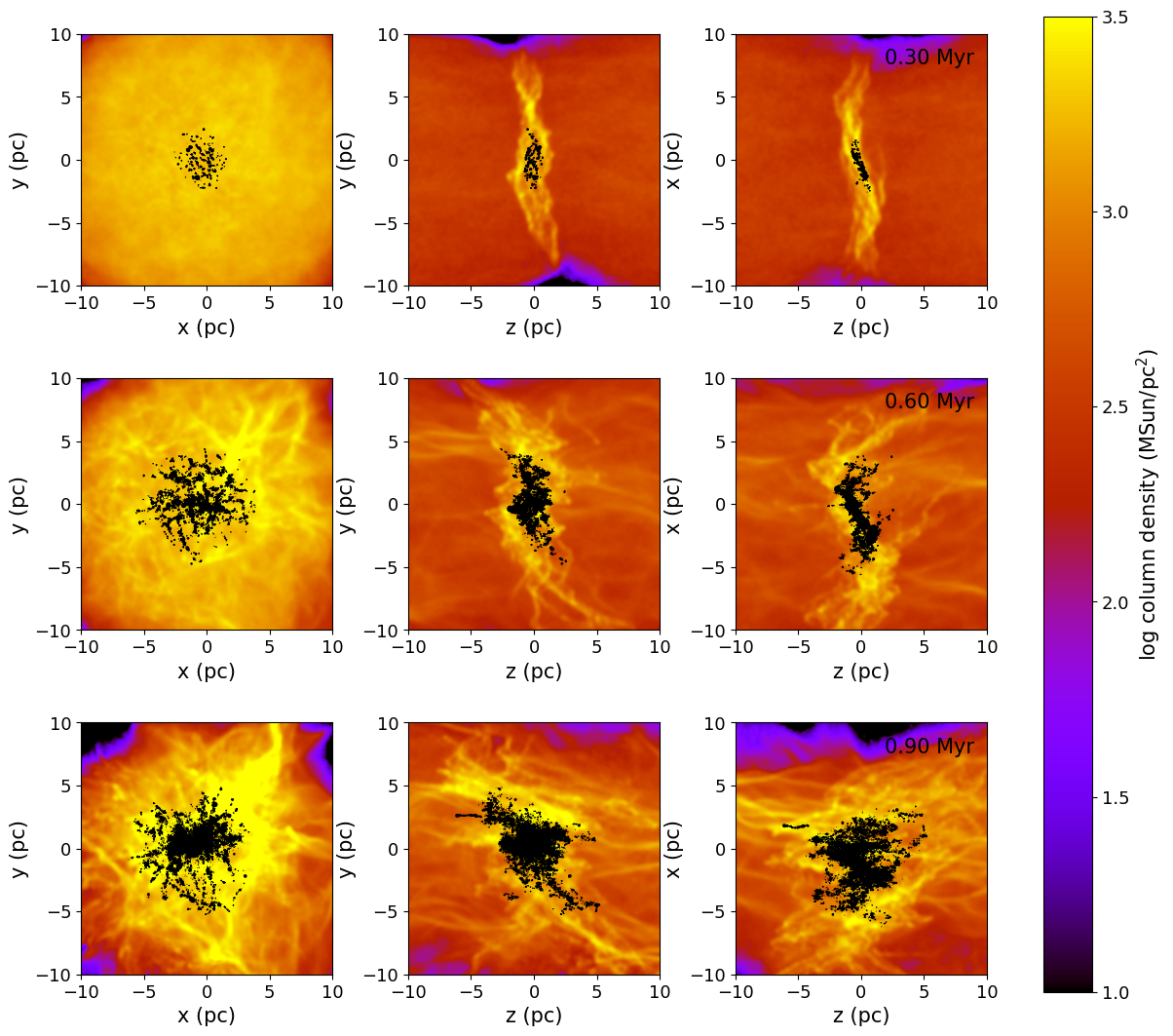}
   \caption{Figure showing our simulated YMC D1050 at 3 different stages of evolution and from different 2D perspectives.
      The cluster is formed from a z-axis cloud-cloud collision. The cluster is shown at an age of (Top row:) 0.3\,Myr, (Middle row:) 0.6\,Myr and (Bottom row:) 0.9\,Myrs, as seen in the (Left column:) X-Y (Middle column:) Y-Z and (Right column:) X-Z plane. Black dots represent stars and the colour scale denotes gas density.}  \label{Fig_simulation} 
\end{figure*}

A full description of the initial setup for our  simulation is presented in \cite{liow_collision_2020} but we also include details here for completeness.
Initially, two ellipsoidal clouds with minor radii of 7\,pc and major radius of 16\,pc are created and aligned along the major axis. Each cloud has mass of $1\times 10^5$ M$_{\odot}$, which gives an initial cloud density of $2.06 \times 10^{-21}$ g cm$^{-3}$ and a spherical free-fall time of 1.46\,Myr. Each cloud contains half of the $5\times 10^6$ SPH particles used in this simulation. The gas particles are positioned randomly within the gas clouds. To simulate turbulence, we impose two separate initial velocity fields, one on each cloud, which are random, Gaussian and divergence-free. The setup of the clouds is similar to that of \cite{bate_formation_2003}. These turbulent fields are consistent with Burger's supersonic field and Larson's scaling relation, and more details can be found in \cite{liow_collision_2020}. In this simulation, the root-mean-square turbulent velocity is set to about 2.5\,km s$^{-1}$. Lastly, the clouds are given a relative velocity of 25\,km s$^{-1}$ to collide along the major axis. 

We use the isothermal equation of state to model the collision, as gas in molecular clouds is usually isothermal \citep{dobbs_gmc_sf_2014}. The temperature is kept at 20 K. 
Following the prescription by \cite{bate_modelling_1995}, we introduce sink particles to replace the dense gas particles when their density exceeds $\rho_{\rm sink} = 10^{-18}$ g cm$^{-3}$. At the point of creation, each sink has an accretion radius of 0.01\,pc. Sinks are converted to stars using a Kroupa IMF \citep{kroupa_2001}, adopting a mass range for stars of 0.01 to 100 M$_{\odot}$. Stars are positioned randomly within the accretion radius of the sink, and with a velocity dispersion similar to the surrounding gas velocity dispersion. We use the grouped star formation in \texttt{Ekster} \citep{liow_grouped_2021}. 
This is a star formation prescription which first collects the sinks into multiple groups. This allows us to sample massive stars, closer to a full IMF, even if the sinks themselves are low mass. In this model, the grouping parameters used are $d_g=1$\,pc, $v_g=1$\,km/s and $\tau_g = 1.65$\,Myr. 

In Figure~\ref{Fig_simulation} we show panels from all three orientations (XY, YZ, and XZ) at three different times (0.3, 0.6 and 0.9\,Myr) from the simulation. From the YZ (and XZ) direction stars initially form along the interface of the shock, since the clouds are colliding in the direction parallel to the $z$-axis, however in the XY direction they appear more spherically symmetric. Over time, the distribution of stars contracts along the $x-y$ axes and becomes more spherical as viewed in the YZ and XZ orientations.

%~~~~~~~~~~~~~~~~~~~~~~~
\subsection{Cluster Set}

As detailed above, the simulation provides full 6D phase space (velocities and positions in the XYZ plane) data for each star. To explore how observer perspective and stellar sample incompleteness can impact the correct determination of a cluster’s properties, we generate 4 cluster realisations (Table\,\ref{Tab_clusters}) from the simulation.

The first realisation consists of all cluster members with their full 6D phase space data intact, which will be used as a baseline for our analysis.  Second, third and fourth realisations are created from the 2D data in each plane (XY, YZ, XZ) and represent how the cluster would appear if viewed as a 2D object in the XY, YZ and XZ planes. For these realisations all members are included, but only 4D phase space data is available for each star (i.e. 2D velocities and positions in the respective plane). 

For each realisation, 66 snapshots of the cluster are taken at 0.01\,Myr intervals between 0.30-0.96\,Myr.

%#########################################################################

\begin{table}
\caption{List of cluster realisations, phase space parameters and which stars these parameters are available for. \label{Tab_clusters} 
} \centering                                      
\begin{tabular}{c | c | c }          
\hline\hline                        
Name & Phase Space & Availability \\    
\hline                                   
    D1050 & $X, Y, Z, V_{x}, V_{y}, V_{z}$ &  All Members\\ 
    D1050XY & $X, Y, V_{x}, V_{y}$ &  All Members\\      
    D1050YZ & $Y, Z, V_{y}, V_{z}$ &  All Members\\      
    D1050XZ & $X, Z, V_{x}, V_{z}$ &  All Members\\      
\hline                                             
\end{tabular}
\end{table}

\section{Method}

We use the following tools to identify features and characterise behaviours in the spatial and velocity parameter spaces of the cluster.

\subsection{INDICATE}\label{sect_indicate}
We quantify stellar member's degree of association using the 2+D INDICATE\footnote{\url{https://github.com/abuckner89/INDICATE}} tool introduced by \citet{2019A&A...622A.184B}. The tool works by examining the local stellar spatial concentration around each star, comparing this to the expected value if the star were not clustered, then assigning an index value to each star accordingly. Advantageously, INDICATE is robust against edge effects, outliers, and there is no dependence between the index cluster shape, size or stellar density (see \citealt{2019A&A...622A.184B} for a discussion).

Index values are derived as follows. An evenly spaced uniform (i.e. definitively non-clustered) control distribution is generated across the parameter space of the cluster with the same number of members as the cluster. For every star $j$ the mean Euclidean distance, $\bar{r}$, to its $5^{th}$ nearest neighbour $i$ in the control is determined in n-dimensional space i.e.

\begin{equation}
\\	r_j^2=\sum_{d=1}^{n}(j_{d}-i^{con}_{d})^{2}
\end{equation}

thus, 

\begin{equation}
\\	\bar{r} = \frac{1}{N_{tot}} (\sum_{j=1}^{N_{tot}} r_j)
\end{equation}

\begin{figure}
\centering
   \includegraphics[width=0.42\textwidth]{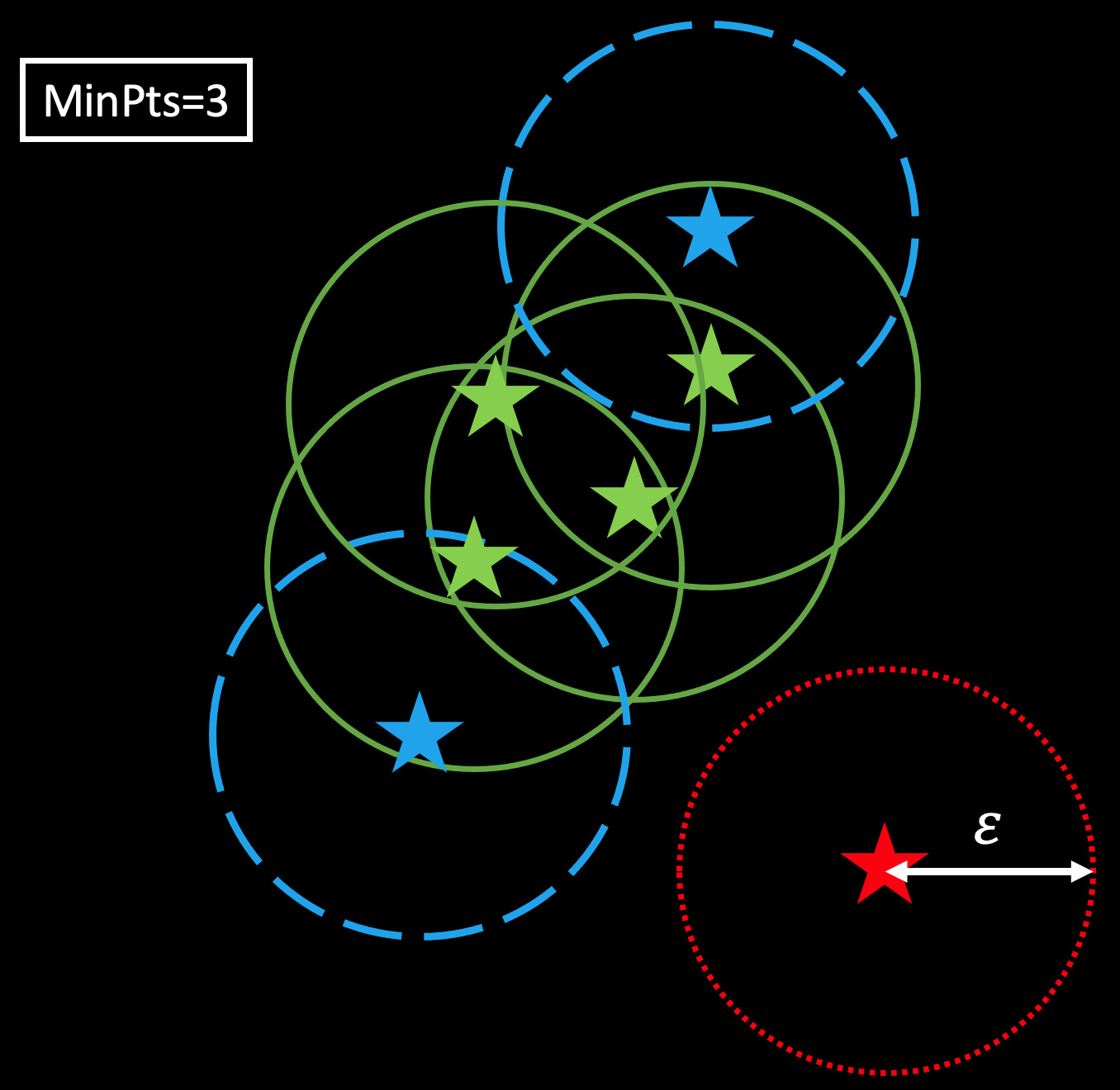}
   \caption{Visual demonstration of how the DBSCAN algorithm works. For each star, the area within radius $\epsilon$ is searched for neighbours. If the number of neighbours is greater than MinPts-1 the star is identified as a `core' star of the subcluster (green in plot). If a star is within the search radius of a core star and (i) has more than MinPts-1 neighbours it is also designed a core star, (ii) has less than MinPts-1 neighbours it is designated a `border' star (blue in plot). Else, the star is identified as `noise' (red in plot).}  \label{Fig_dbscan} 
\end{figure}

where $N_{tot}$ is the total number of cluster members, $n$ the number of dimensions of the parameter space, and $x$ the component of position along dimension $v$. 

The index of star $j$ is then calculated as the ratio of the actual and expected number of nearest neighbours to star $j$, such that

	\begin{equation} \label{eq_I}
		\\ I_{5,j}= \frac{N_{\bar{r}}}{5}
	\end{equation}

where $N_{\bar{r}}$ is the actual number of nearest neighbours to star $j$ in the cluster, and $I_{5,j}$ is the unit-less index of the star with a value range of $0 \le I_{5,j} \le \frac{N_{tot}-1}{5}$. The higher the value of $I_{5,j}$ the greater the star’s degree of association. 

Calibration with random distributions is performed to identify index values which denote a star is spatially clustered (rather than randomly distributed). For a cluster, the mean index value, $\bar{I_5}^{random}$, of 100 realisations of a random distribution generated across the parameter space with $N_{tot}$ points is determined. Cluster star $j$ is then spatially clustered if it is greater than the significance threshold, defined as

	\begin{equation}\label{eq_IjIsig}
	\\ I_{5,\,j}> I_{sig} \text{\,,\,\,\,\,\,where\,\,\,\,\,} I_{sig}=\bar{I_5}^{random}+3\sigma
	\end{equation}

where $\sigma$ is the standard deviation of $\bar{I_5}^{random}$. Using this criteria, $99.7\%$ of stars that are distributed in a spatially random configuration will have an index value below the significance threshold ($I_{5,j}<I_{sig}$). 

An advantage of INDICATE is that in addition to exploring the spatial behaviours of the general populous, it can reliably detect and quantify signatures of Type I and Type II mass segregation in clusters \citep{george_indicate_paper} defined respectively as the concentration of (i) high mass stars typically at the centre of a cluster and ii) lower-mass stars  around high-mass (which are not necessarily concentrated together). 
This is because the index is calculated for every star, and its value represents the strength of the stellar concentration in its local neighbourhood. Therefore, if INDICATE is run on the entire population the values of the high-mass stars by definition provide a measure of (i). If INDICATE is run on just the massive population of the cluster, the index represents the relative concentration strength of high-mass stars with each other, thus providing a measure of (ii). 

%#############################################
% INDEX FIGURES

\begin{figure*}
\centering
   \includegraphics[width=0.5\textwidth]{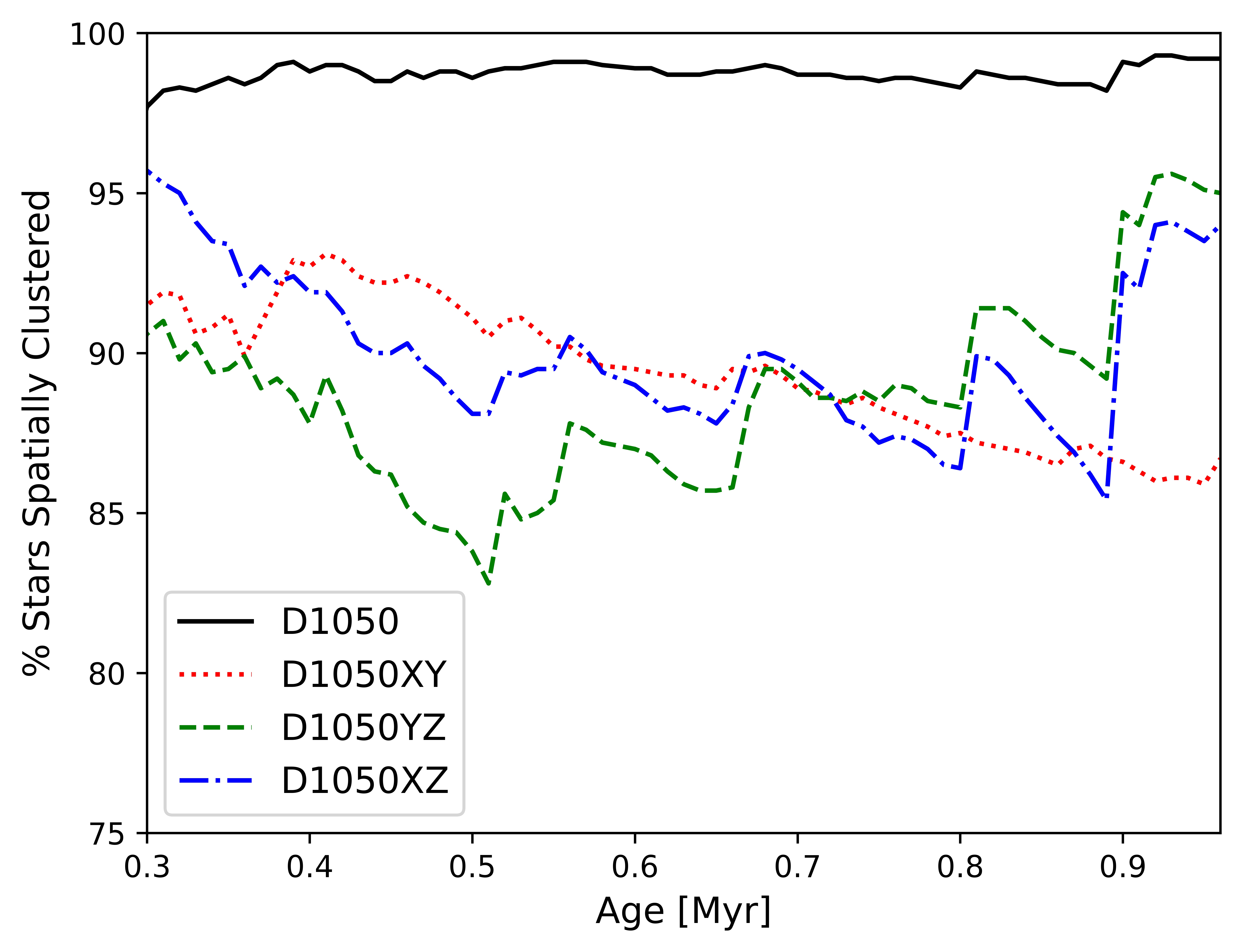}\hfill
   \includegraphics[width=0.5\textwidth]{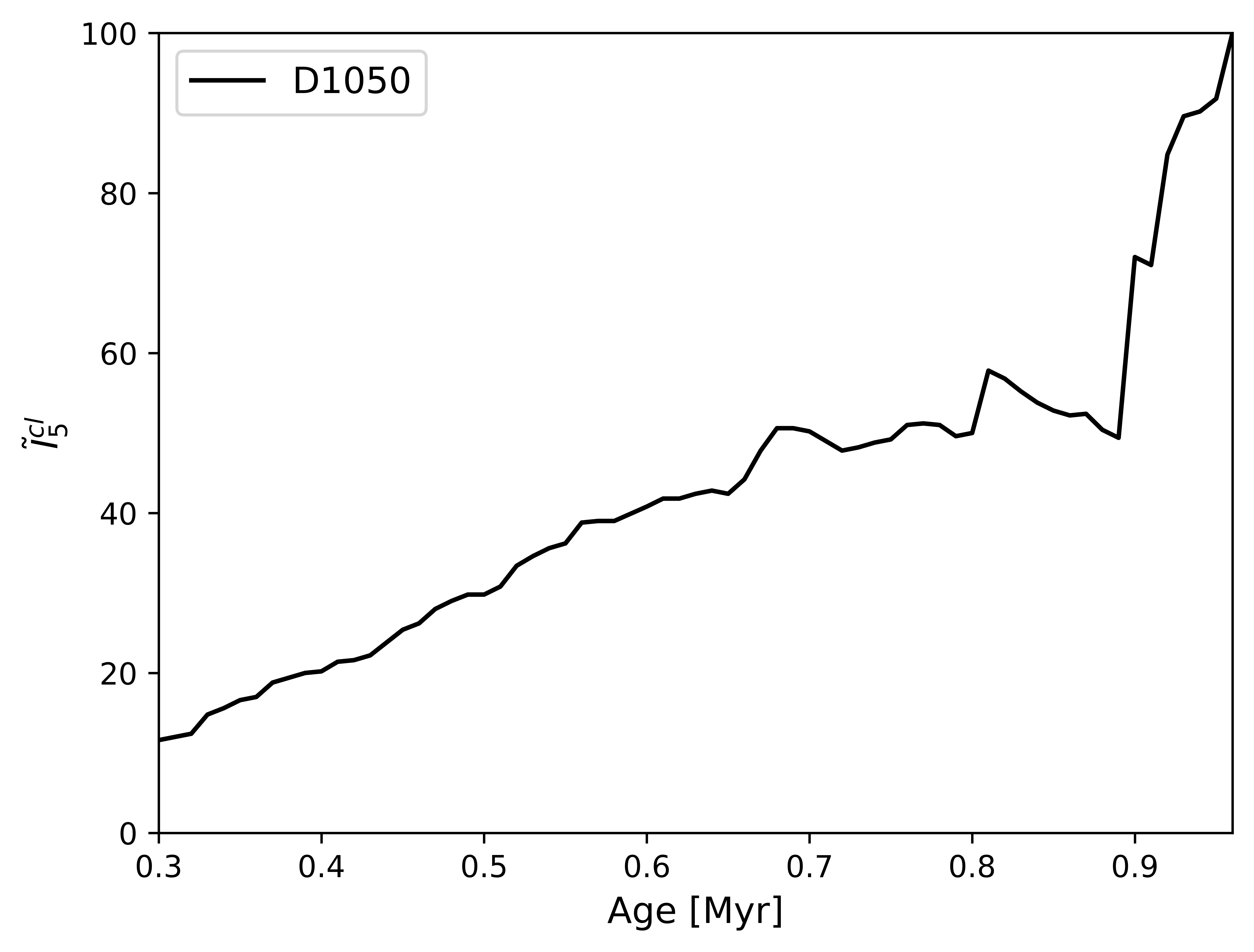}\hfill
   \includegraphics[width=0.5\textwidth]{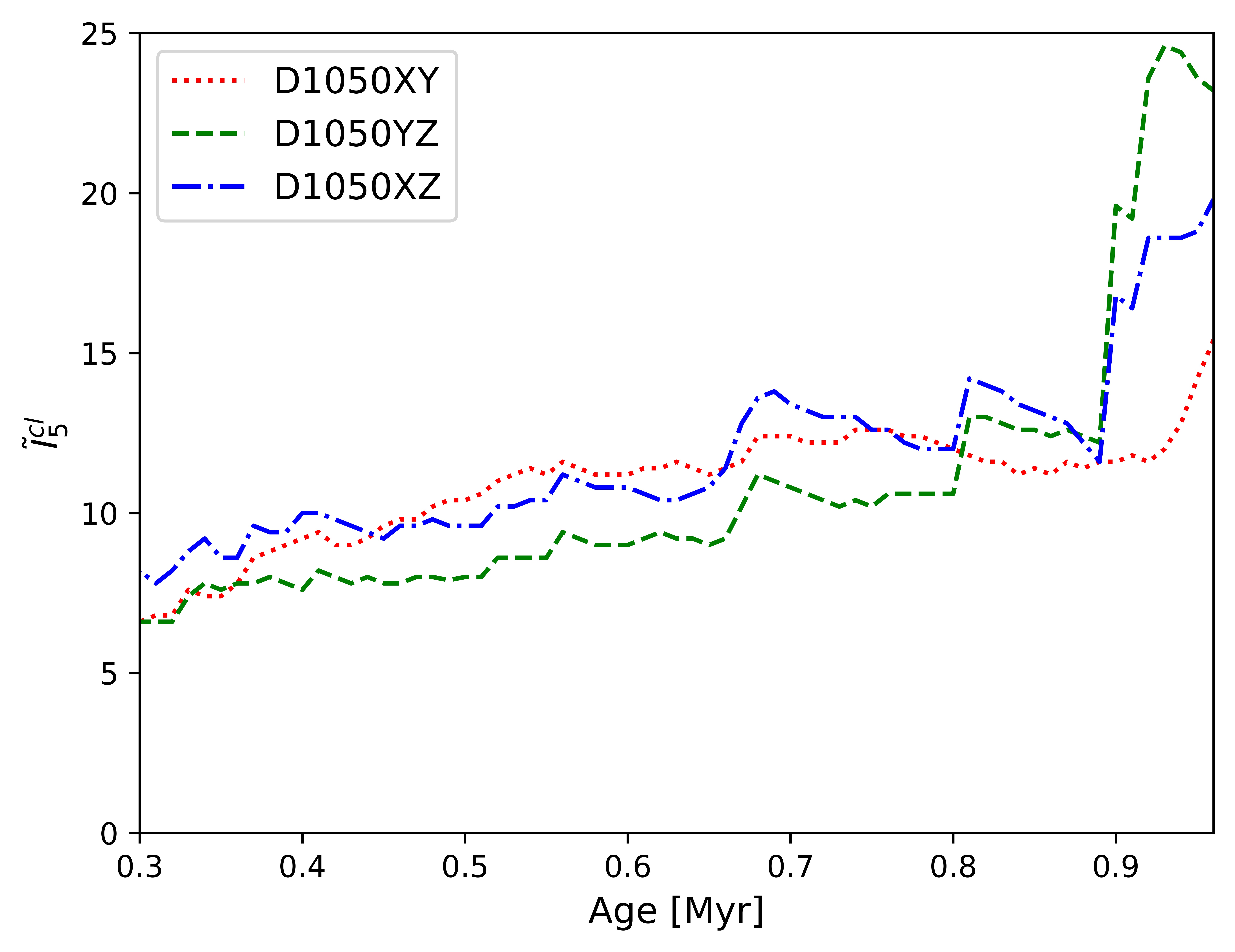}
   \caption{This figure shows the results of the INDICATE analysis as a function of cluster age in 2D and 3D. (Left:) Percentage of members found to be spatially clustered, which is consistently underestimated in the 2D perspectives. Median index value of stars identified as spatially clustered in (Middle:) 3D and (Right:) each of the 2D perspectives. Index values derived for the 2D are underestimated with respect to 3D but trends are preserved.}  \label{Fig_index_all} 
\end{figure*}

%#############################################

%~~~~~~~~~~~
\subsection{DBSCAN}

We employ the Density-Based Spatial Clustering of Applications with Noise (DBSCAN; \citealt{Ester96adensity-based}) algorithm to detect discrete subclusters of stars in the datasets, chosen for this work as it tends to be favoured by observers to quantify structure in cluster regions (\citealt{2014RAA....14..159G},  \citealt{2017A&C....18....1B}, \citealt{2019ASPC..523...87J}, \citealt{ 2019A&A...628A.123Z}, \citealt{2020BAAA..61R...80A}, \citealt{2021PASJ...73..652G}, \citealt{2021A&A...647A..14G}, \citealt{ 2021ApJS..254...20L}, \citealt{2021PASJ...73..652G}) perhaps because it offers several advantages over other well-established clustering algorithms. First, it does not require a priori knowledge of the number of clusters in a region as this is determined by the algorithm itself. This is particularly apt in star formation regions where the structure is often complex and the number of subclusters is not easily ascertained by visual inspection. Second, the shape of found subclusters is not necessarily spatially symmetric; again advantageous in regions of complex structure. Third, not all stars are assigned to a subcluster and can be identified as ‘noise’ (i.e. part of a wider distributed population), and DBSCAN is also robust against outliers. Finally, the algorithm’s output distinguishes between ‘core’ (central) and ‘border’ (edge) stars in each subcluster which can be useful if, for example, further kinematic study is later undertaken.

In this context the definition of ‘subcluster’ is a discrete spatial grouping of stars at a density greater than random expectation, not a definitively gravitationally bound grouping (though this may be true as well). Indeed, it is possible for DBSCAN to find spatial groupings of stars which subsequent kinematic study will show are not a physical grouping. However the goal of this work is to investigate the influence of 2D orientation on perceived properties of the cluster, using tools typically used by observers, rather than determine how well those tools produce physically motivated clusters or subclusters.
The impact of sample incompleteness on perceived cluster properties will be subject of Paper II in this series.

DBSCAN works as follows. For each star, a search is performed for neighbours within radius $\epsilon$. If a star has a total number of neighbours (self-included) greater than or equal to a user-set minimum, MinPts, the star is designated as part of the core of the subcluster. If a star is within the search radius of a core star but has less neighbours than the minimum, it is designated a `border' star of the subcluster. If a star fails to meet the criteria to be either a core or border star, it is designated ‘noise’ (see Figure\,\ref{Fig_dbscan}). The values of MinPts and $\epsilon$ are free parameters set by the user. As a general rule MinPts should be set at $\ge D+1$ where $D$ is the dimension of the dataset (i.e. $\ge3$ for 2D, and $\ge4$ for 3D, datasets). In this work we follow the prescription of \citet{jrg1998densitybased} to set MinPts$=2\times\,D$. To derive the optimal value of $\epsilon$ we:

\begin{enumerate}
\item determine the distance to the MinPts$^{th}$ nearest neighbour for each star (k-distance)
\item order the distances from shortest to longest, then number the stars
\item plot star number vs. k-distance
\item $\epsilon$ is equal to the k-distance which corresponds to the maximum curvature of the line, found using the \texttt{KneeLocator} class of the \texttt{kneed} Python package.
\end{enumerate}

%~~~~~~~~~~~
\subsection{Q}\label{sect_q}

To distinguish between a fractal, random and radial density gradient spatial configuration of the cluster, we use the $\mathcal{Q}$ parameter by \citet{2004MNRAS.348..589C}. This parameter is defined as the ratio between the Normalised Mean Edge Length $\bar{m}$ and Normalised Correlation Length $\bar{s}$ of cluster members i.e.
\begin{equation}
\\ \mathcal{Q}=\frac{\bar{m}}{\bar{s}}
\end{equation}
where $\bar{s}$ is the mean Euclidean distance between members divided by the cluster’s radius; and $\bar{m}$ is calculated from the Minimum Spanning Tree of members positions as the mean branch length divided by a normalisation factor of

\begin{equation}
\\ \sqrt{\frac{A}{N_{tot}}}    \text{\;\;\;\;\;\;\;\;(for 2D)}
\end{equation}
\begin{equation}
\\ \sqrt[3]{\frac{V}{N_{tot}}}        \text{\;\;\;\;\;\;\;\;(for 3D)}
\end{equation}

with $A$ and $V$ the area and volume of the cluster respectively.

When applied to 2D positional data, values of $\mathcal{Q}<0.8$, $\mathcal{Q}\approx0.8$ and $\mathcal{Q}>0.8$ indicate a cluster has a fractal, random and radial density gradient configuration respectively \citep{2004MNRAS.348..589C}. When applied to 3D positional data, the boundary values distinguishing fractal, random and radial density gradient configurations are slightly lowered at $\mathcal{Q}<0.7$, $\mathcal{Q}\approx0.7$ and $\mathcal{Q}>0.7$ respectively \citep{2009MNRAS.400.1427C}.

%~~~~~~~~~~~
\subsection{Velocity Analysis}

We determine if the cluster is expanding or contracting through inspection of the directional velocities of stellar members. If expanding, the majority of stellar velocities, $v_{*}$, should be directed outwards from the cluster centre; and if contracting, directed inwards. Therefore, following the prescription of \citet{2019ApJ...870...32K}, we express the stellar velocities of all stars in terms of their outward velocities, $v^{*}_{out}$, w.r.t the cluster centre 

\begin{equation}\label{Eq_vout}
 \\ v^{*}_{out}=\vec{v}_{*}\cdot \hat{r}^{n}
\end{equation}

where $\hat{r}^{n}$ is the n-dimensional outward component of velocity

\begin{equation}
 \\ \hat{r}^{n}=(r_{1},r_{2},….,r_{n})
\end{equation}

and derive the median outward velocity, $\tilde{v}_{out}$, for the cluster. Negative values of $\tilde{v}_{out}$ denote cluster contraction, and positive values expansion.

\subsection{The O-ring function}

The O-ring function measures the mean surface density of objects at a distance $r$ from an object, averaged over all objects.
Practically this has to be calculated over an annulus, whose area we define as $A(r)$ and half width as $q$.
This leads to the definition we will use in this paper,
\begin{equation}
\\    \mathcal{O}(r)={\frac{2}{N_{\rm tot} A(r)}}{\sum_{i=1}^{N_{\rm tot}}}{\sum_{j>i}^{{N}_{\rm tot}}}I_{r}(i,j).
    \label{eqn_oring}
\end{equation}
Here $I_{r}(i,j)$ is a selection function which takes the value one if the distance between star particles $i$ and $j$ lies between  $r-q$ and $r+q$, otherwise it is zero. 

In some form this function has been applied to a wide range of problems. In ecology it is used to test the spatial scales at which distributions deviate from complete spatial randomness by comparison with simulation \citep[e.g.][]{2004Oikos...104...209W}.
In an astronomical context it has been demonstrated to be a useful test of whether a distribution of sources is random \citep{2019MNRAS.487..887R} or in an adapted form whether a distribution follows a particular generating function \citep{2021MNRAS.507.1904R}.

In this paper however, we will use Equation \ref{eqn_oring} in a similar way to \cite{1995MNRAS.272..213L} and \cite{1997ApJ...482L..81S} to examine the spatial scales on which clustering occurs, rather than as a test of significance, and hence will refer to the O-ring function, rather than (as used in some contexts) the O-ring statistic. 
Changes in gradient in the O-ring function indicate where the distribution has a characteristic length scale, whereas a straight line indicates a power law (and therefore scale-free) relationship between density and distance.
To compare the results of the simulation in two and three dimensions a three-dimensional version of the O-ring function is needed. We achieve this by the straightforward translation of Equation \ref{eqn_oring} into three dimensions by using the the three dimensional distance for $r$, and replacing $A(r)$ by the volume enclosed between spheres of radius $r-q$ and $r+q$.

Alternatively we could have used the two-point correlation function widely used in cosmology, but as explained in \citep{2019MNRAS.487..887R} and 
\cite{1995MNRAS.272..213L} the O-ring function is more intuitive in this context. 

%~~~~~~~~~~~~~~~~~~~~~~~~~~~~~~~~~~~~~~~~~~~~~~~~~~~~~~

\section{Results}

\begin{figure}
\centering
   \includegraphics[width=0.47\textwidth]{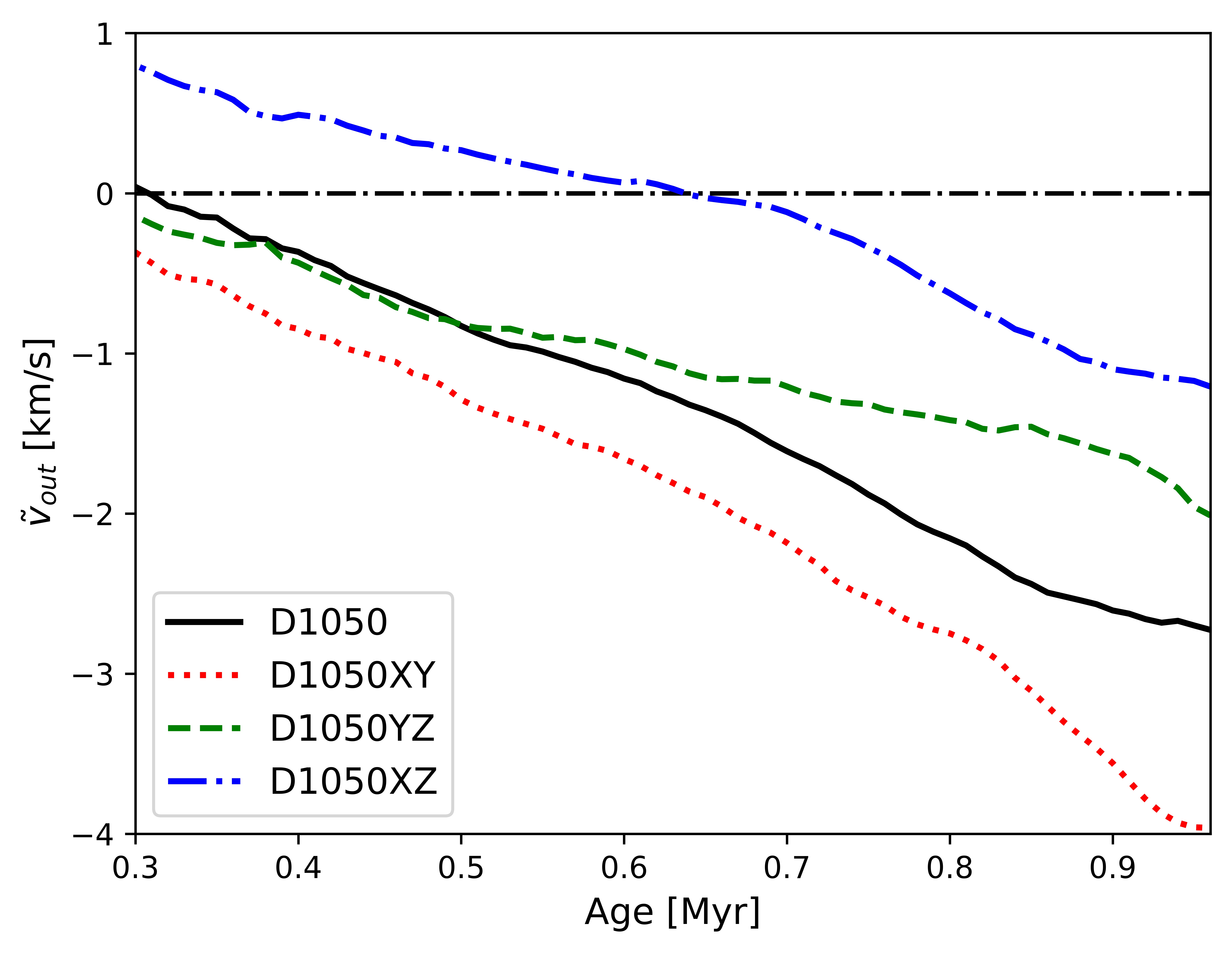}
    \includegraphics[width=0.49\textwidth]{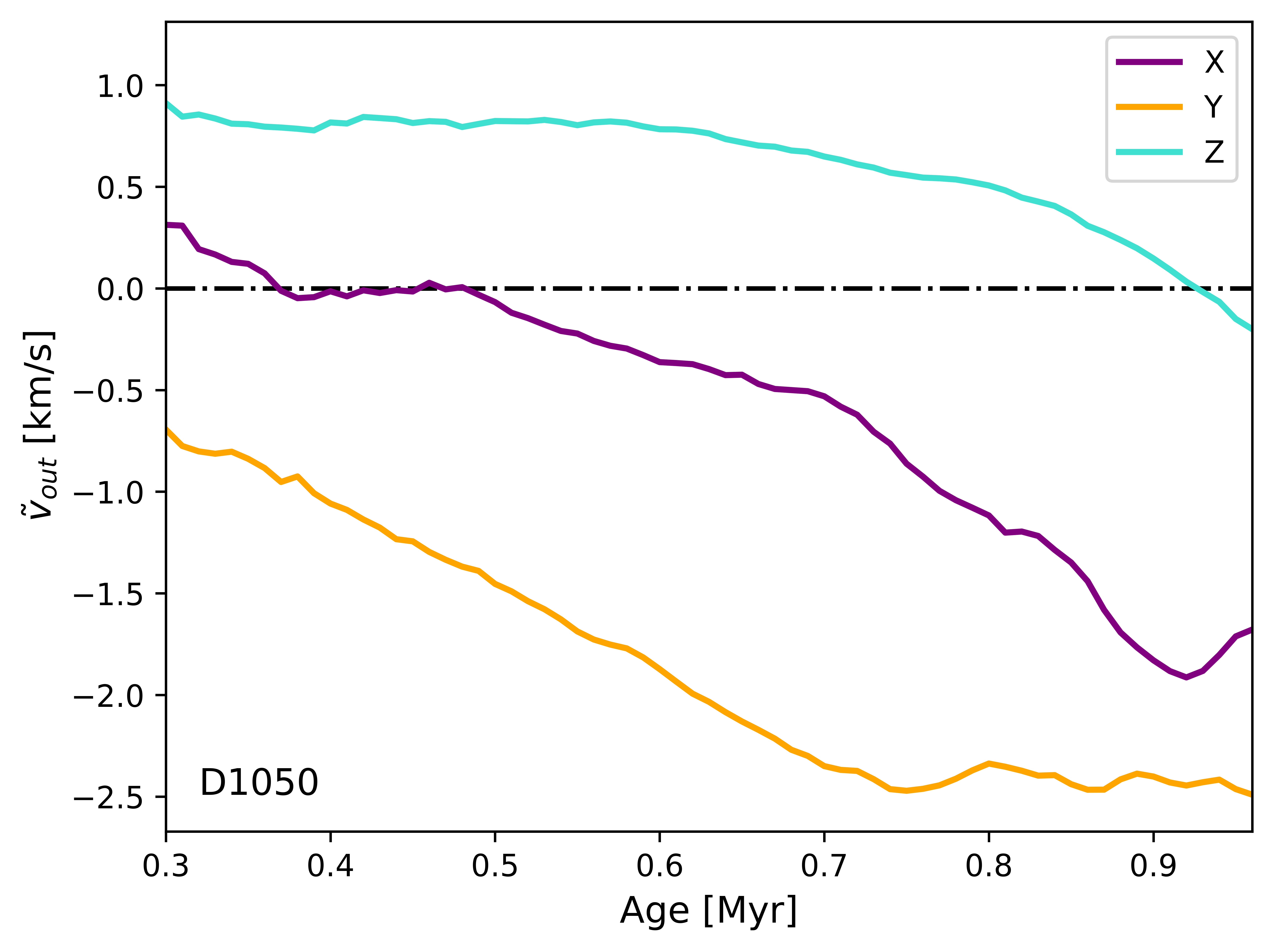}
   \caption{This figure shows the (Top:) expansion velocity from 0.30-0.96\,Myrs for  D1050, D1050XY, D1050YZ and D1050XZ; and (Bottom:) mean 1D directional velocity along the X, Y and Z axes. Positive values indicate expansion, negative contraction, and zero a static state (marked by the horizontal black dot-dash line).}  \label{Fig_vout} 
\end{figure}

In this section we discuss the cluster's spatial and kinematic properties as determined from the full 6D and (2+2)D data to assess the impact of perspective effects and whether the evolution appears qualitatively similar irrespective of these. Comparative visualisations of the results discussed below are provided in Figures\,\ref{Fig_index_all}-\ref{Fig_fcorrect}. We first provide a description of the properties of the cluster using the 6D data in Section\,\ref{results_3D}, before comparing those derived from the limited (2+2)D data in Section\,\ref{results_2D}.

\subsection{Overview of D1050}\label{results_3D}

Figure\,\ref{Fig_index_all} shows the fraction of stars clustered (left) and the median index value of clustered stars (centre, right) versus time, as given by INDICATE. As shown in Figure\,\ref{Fig_index_all}, for the full 6D case there is a gradual increase in both the proportion of stars spatially clustered, and the degree of their associations, as the system evolves.  In the first snapshot at 0.30\,Myr, $95.7\%$ of members are clustered, with a median index value for the population of $\tilde{I}_5=10.2$; but by 0.96\,Myr $99.2\%$ of members are spatially clustered and $\tilde{I}_5=100.0$. This suggests the cluster initially forms with the majority of its population in relatively loose spatial concentration(s), then shortly thereafter the spatial configuration rapidly and significantly changes such that a higher proportion of the population is in spatial concentration(s) that typically increase their degree of association by a factor of $\sim§10$. We perform a two sample Kolmogorov–Smirnov Test (2sKST) with a strict significance boundary of $p < 0.01$ to test whether the distributions are distinct and our assertion is correct. Retrieving $p<<0.01$, we reject the null hypothesis: the spatial behaviours of stars at 0.30\,Myr and 0.96\,Myr are distinct and the spatial configuration of the cluster does indeed significantly reconfigure. 

\begin{figure*}
\centering
   \includegraphics[width=0.49\textwidth]{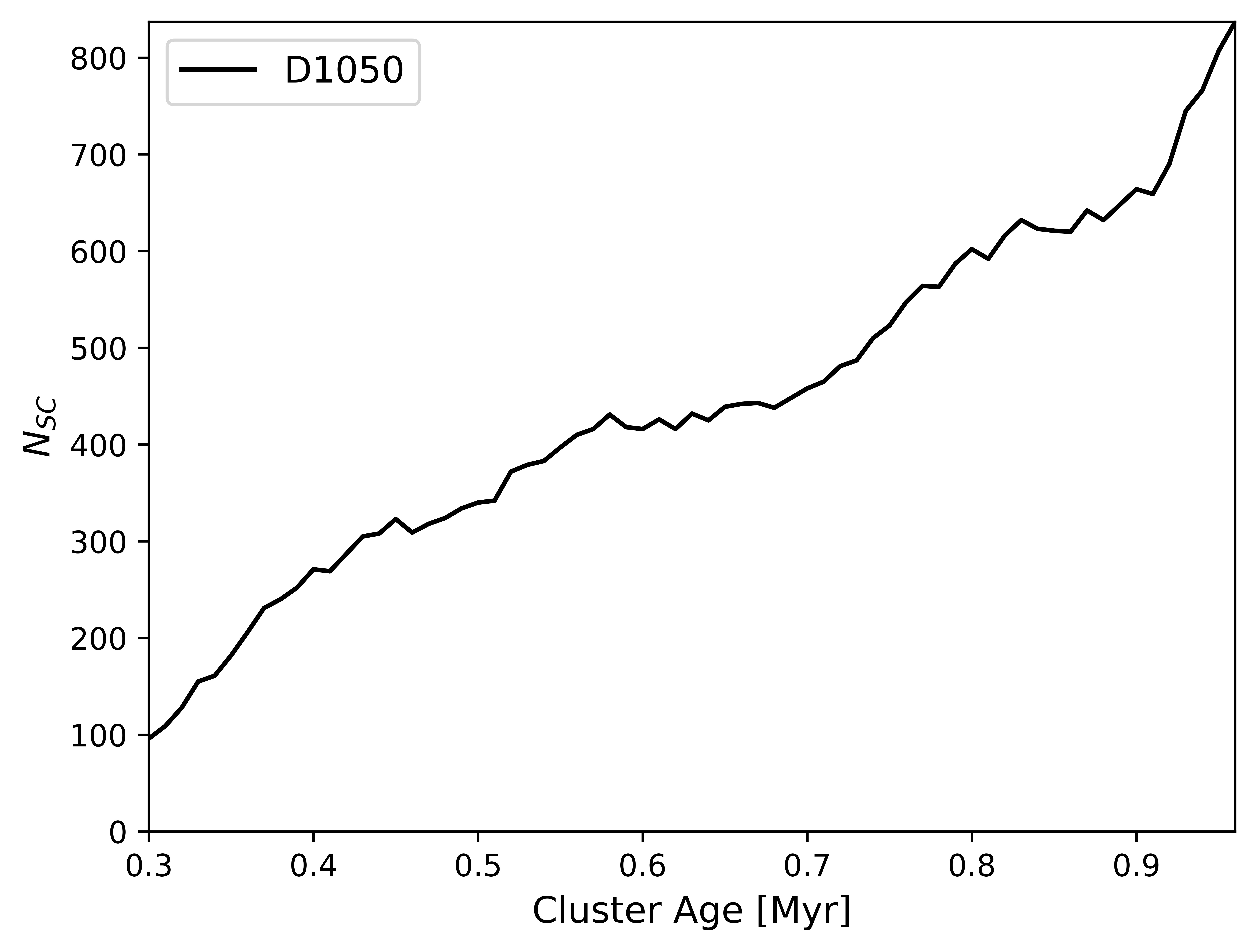}\hfill
   \includegraphics[width=0.5\textwidth]{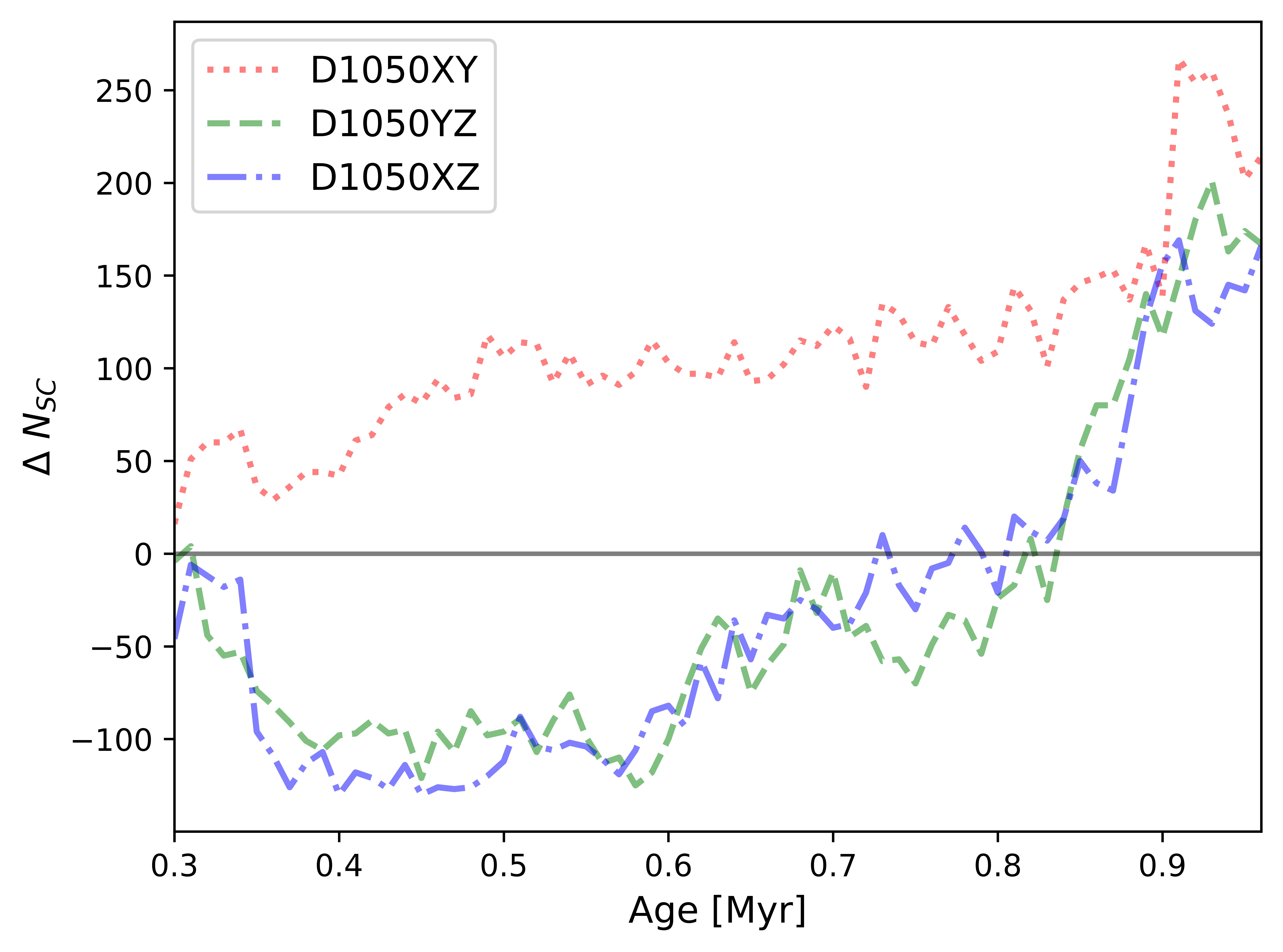}\hfill
      \includegraphics[width=0.5\textwidth]{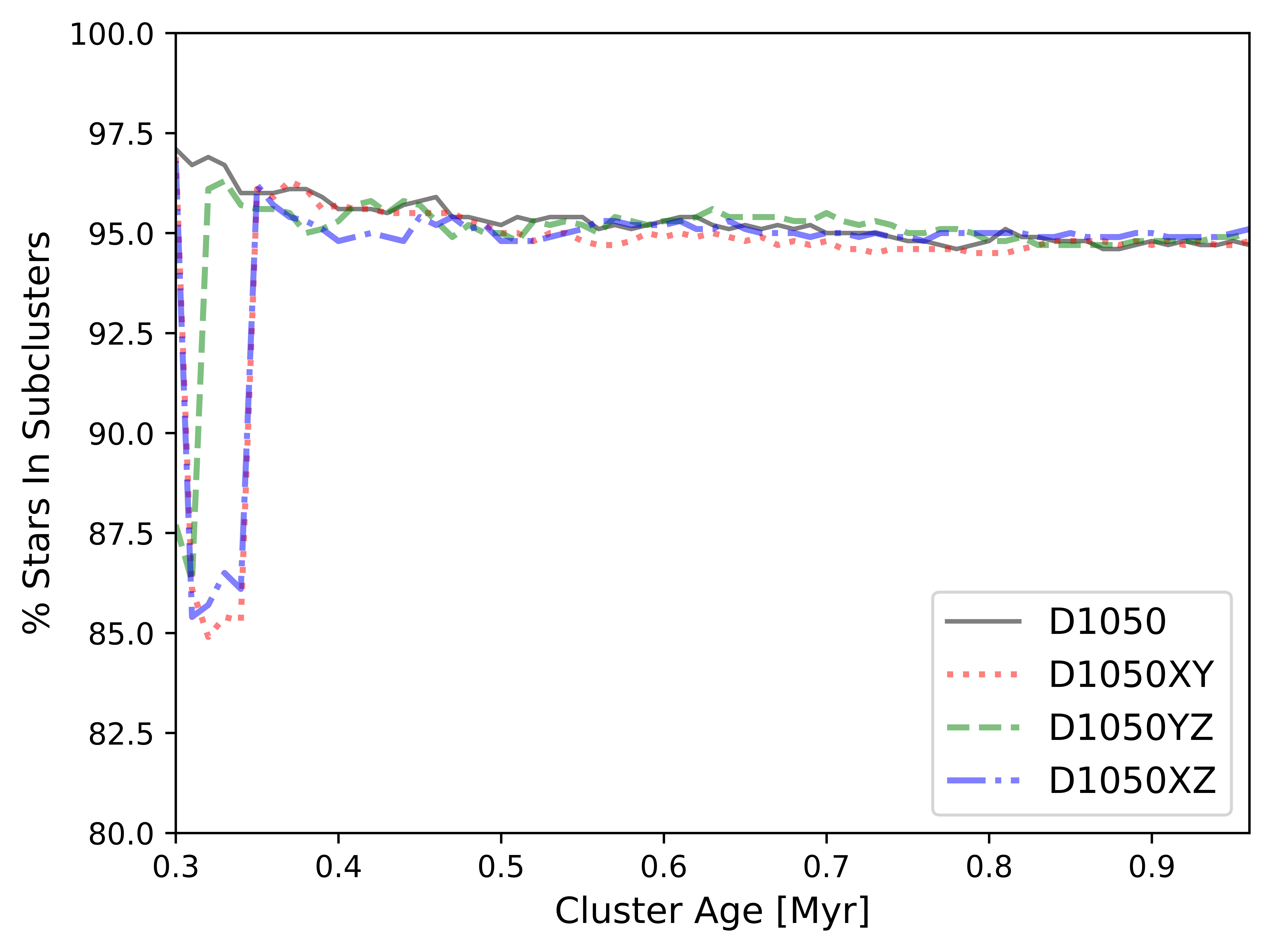}
   \caption{This figure shows the results of the DBSCAN analysis as a function of cluster age in 2D and 3D. (Left:) The number of subclusters found using the full 3D spatial information and (Middle:) in 2D with respect to 3D where $\Delta N_{SC} =0$ means the same number of subclusters were found in the 2D and 3D data, and $\Delta N_{SC}>0$ means the number of subclusters is overestimated in 2D. (Right:) The percentage of the population found to be in subclusters.}  \label{Fig_ncl} 
\end{figure*}

\begin{figure}
\includegraphics[width=0.49\textwidth]{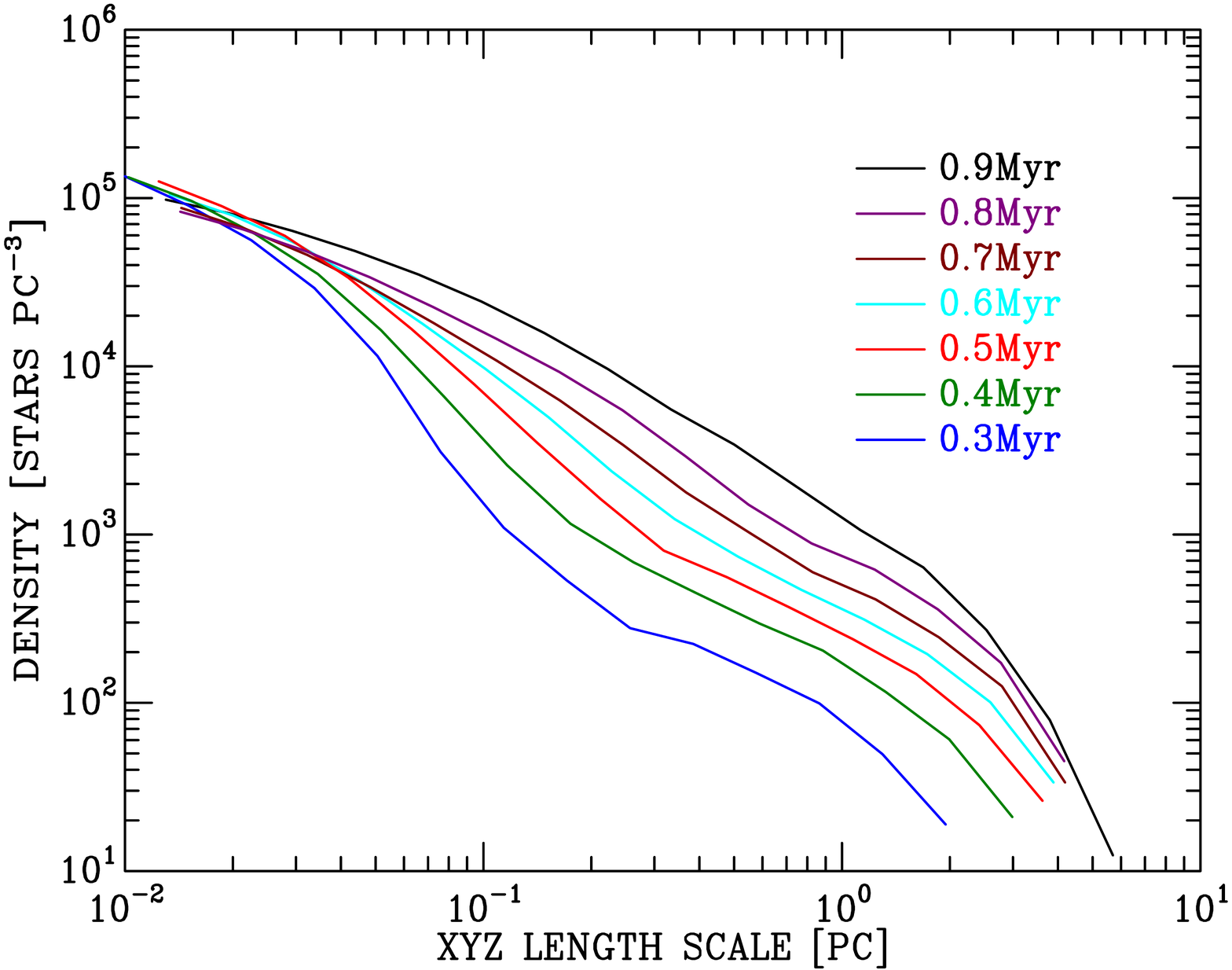}
\includegraphics[width=0.49\textwidth]{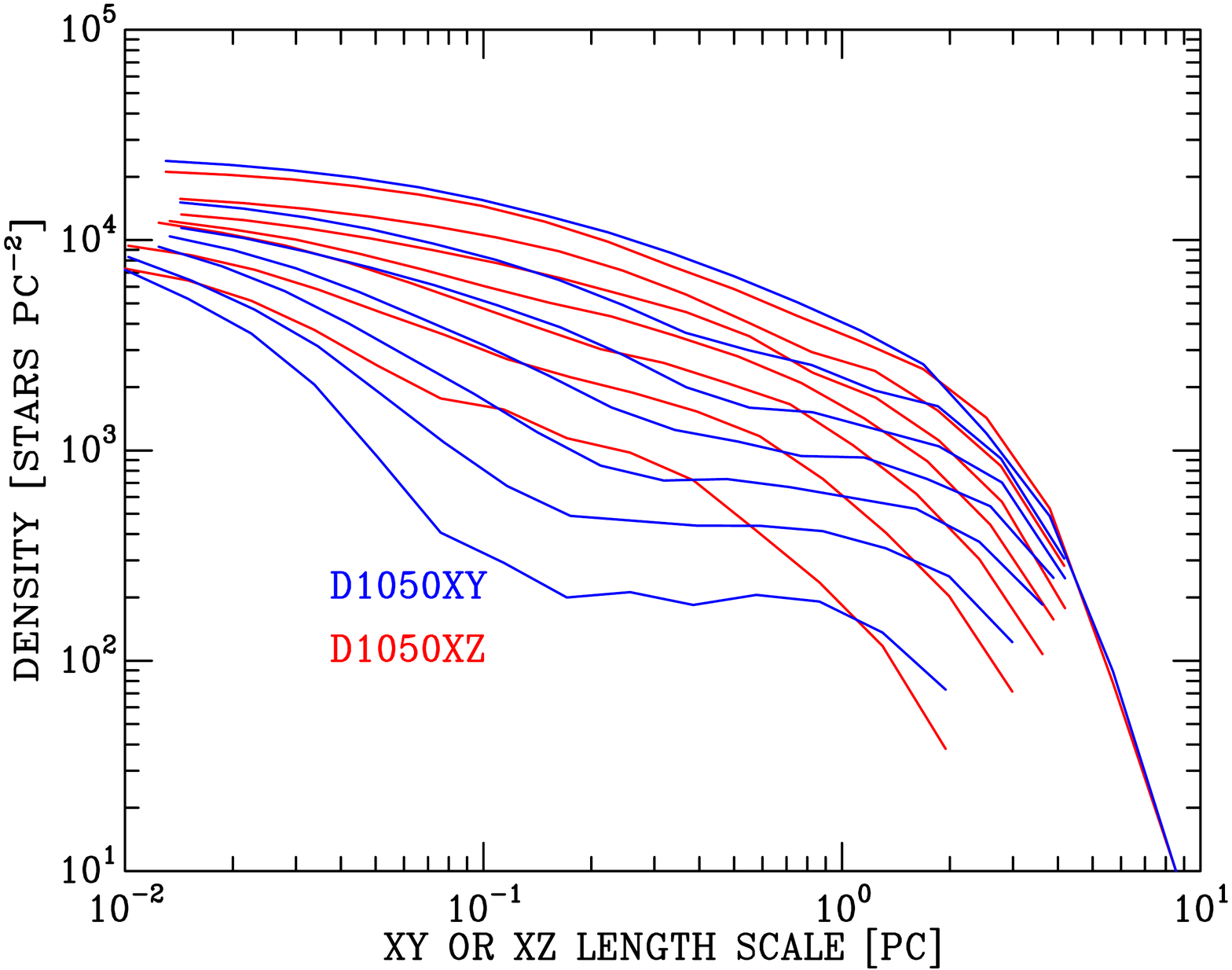}  
   \caption{
   The density of stars as a function of radius derived from the O-ring function in (Top:) 3D and (Bottom) 2D.
   The O-ring functions are shown from 0.3\,Myr (lowest curve on right-hand side) to 0.9\,Myr (highest curve on right-hand side) in 0.1\,Myr steps. 
   } \label{Fig_oring_comb}
\end{figure}

In regards to the nature of the spatial reconfiguration, we find that the evolution of the cluster is not driving it towards a centrally concentrated structure, but rather an increasingly dense clumpy one. As seen in  Figure\,\ref{Fig_simulation}, at later times the central $\sim 2$ pc region is more densely filled with stars, which INDICATE shows are in clumpy structures. The increase in  stars in the centre is the result of both ongoing star formation and a 
collective inward stellar motion. This motion is confirmed by our velocity analysis (Fig.\,\ref{Fig_vout}) and shows a sustained, accelerating, contraction of the system.
Evidence for a heavily substructured configuration is provided by the number of distinct subclusters identified, which increases from 96 at 0.30\,Myrs to 837 at 0.96\,Myrs (Figure\,\ref{Fig_ncl}) while the fraction of the population within these subclusters slightly decreases from $97\%$ to $95\%$. Moreover, the cluster's $\mathcal{Q}$ value undergoes a steady increase but remains firmly $<<0.7$ at all age steps, that is, the density gradient remains in a fractal configuration as the cluster evolves. The O-ring functions in the upper panel of Figure \ref{Fig_oring_comb} confirm the growing density of the sub-clusters.
In a thought experiment where stars are born at random positions in the simulation and do not move far from their birthplaces we would expect to see a uniform rise in the density, instead of which the O-ring functions show no rise at the shortest length scales, and the largest growth in density at about 0.2\,pc from other stars.
Stars may be more likely to be born around 0.2\,pc from other stars as this length scale is similar to the Jeans length of the region, ( $\approx$ 0.38\,pc), which means that gravitational instability is likely to occur on these length scales.

If Type I mass segregation is present the majority of massive stars should be spatially clustered with other high mass members. As shown in Figure\,\ref{Fig_MS_i} we find mass segregation to be present at all cluster ages. 

As the cluster evolves there is an increase in high mass stars spatially clustered together from $63.0\%$ to $94.5\%$, and the tightness of the clustering increases by a factor of $\sim21$. The prominent increases at $\sim 0.9$\,Myrs in both the number of stars clustered, and the tightness of this clustering, coincide with the onset of gravitationally driven star formation in the cluster. Examination of the positions of massive stars suggest they are forming in relative isolation towards the outer edges of the (high mass) stellar distribution, even after the onset of gravitationally driven star formation, then dynamical evolution is driving them to spatially cluster together along the collision axis in an asymmetric clumpy amalgamation ($Q<0.7$). 

If Type II mass segregation is present the majority of high mass stars should be clustered in stellar concentrations with other members, and the intensity of these concentrations should be significantly greater than those of the low/intermediate mass (LIM) population. As shown in Figure\,\ref{Fig_MS_ii} we find that initially most, but proportionally fewer, massive stars are in stellar concentrations than LIM stars and that typically the intensity is the same for both populations (i.e. no mass segregation is present). After $\sim\,0.75$\,Myrs the majority of high mass stars are still clustered but the proportion is now equivalent to the LIM population, and the typical intensity for massive stars is significantly greater (confirmed by a 2sKST). This suggests that high mass stars are initially more likely to form in isolated areas than their lower mass counterparts, but generally form in areas of comparable stellar concentrations, then as the cluster evolves most are mass segregated but a small number make up part of the dispersed population. Again, the notable spike in the tightness of clustering at $\sim 0.9$\,Myrs is consistent with the onset of gravitationally driven star formation in the cluster \citep[see the Mach 50 collision star formation rate line in first row, third column of Figure 5 in][]{liow_collision_2020}.

As expected the above analysis using the full 6D data agrees with the evolution visually observed to occur in the simulation: stars initially form along the interface of the shock and the clustering is dominated by the initial positions of the star particles as spatially separate sinks (Figure\,\ref{Fig_simulation} top left panel) are formed and converted into discrete clusters. 
Over time gravitational collapse dominates and the star formation occurs preferentially in one main cluster centred about the collision axis, with the result that rather than seeing discrete areas of star formation, stars cover most of the central $\sim 2$ pc radius area (Figure\,\ref{Fig_simulation} middle and lower panels). The large increase in the degree of stellar association that occurs at $\sim 0.9$\,Myr in Fig.\,\ref{Fig_index_all} potentially marks the increase in star formation across the region due to gravitational collapse rather than the previously dominant collision-based formation.\\

%#############################################

%#############################################

\subsection{2D Analysis}\label{results_2D}
\subsubsection{Clustering}

Viewed from different orientations, the perceived spatial behaviour of the stars varies. As seen in Figure\,\ref{Fig_simulation}, D1050XY has a notably different spatial symmetry compared to the other orientations being disk-like in appearance rather than elongated, and the distribution of the stars (especially at early times) is quite different also. The percentage of stars which are spatially clustered progressively decreases from 0.41\,Myrs in D1050XY, whilst for D1050YZ and D1050XZ there is an initial decrease which is followed by an increase at 0.53 and 0.89\,Myrs respectively (Figure\,\ref{Fig_index_all}). All have lower percentages of clustered stars compared to D1050, the 3D results. Despite this, all three orientations display broadly similar spatial behaviour as the system evolves. There is a steady increase in stars' degrees of association and a rapid increase at $\sim 0.9$\,Myr (consistent with the 6D results). Typically stars appear to increase their degree of association by a factor of $\sim$2.5-4, depending on perspective. Similar to the 3D case, 2sKSTs tests confirm a distinct change in the spatial behaviours of stars between 0.30\,Myr and 0.96\,Myr is observed in each orientation and that spatial reconfiguration occurs.

The number of subclusters identified in D1050YZ and D1050XZ is underestimated w.r.t the actual number present until $\sim0.8$\,Myrs, when thereafter the number is overestimated but the proportion of stars found to be in subclusters is consistent with that found in the 3D case (Figure\,\ref{Fig_ncl}). This suggests the increased degree of stellar association of the cluster at this time has resulted in the detection of some real singular subclusters as numerous smaller ones with some of the real members wrongly assigned to the dispersed field. A similar effect is observed in D1050XY where the number of subclusters found is consistently overestimated w.r.t. the number actually present, but after 0.9\,Myrs there is a marked increase in the overestimation and to a degree similar to that of D1050YZ and D1050XZ. At ages $<0.35$\,Myrs the fraction of the population within the found subclusters is underestimated by up to $12\%$ in each orientation. However as shown in the right panel of Figure\,\ref{Fig_ncl} (with the exception of the above discussed deviations) the fraction of stars in subclusters is typically consistent and correct.

We compared the membership of the subclusters in the 2D versus the full 3D spatial data: were stars being correctly grouped together? Figure\,\ref{Fig_fcorrect} left panel shows the fraction of subclusters which are identified with exactly the same members as found in the 3D perspective. Indeed, for all 2D perspectives this fraction is typically small. For D1050XY the fraction of correctly identified subclusters – that is consist exclusively of (and include all) true member stars – is greatest but decreases as the cluster evolves, from a peak of $56.3\%$ at 0.36\,Myrs to just $3.4\%$ by 0.96\,Myrs. For D1050YZ and D1050XZ, the fraction peaks at $<14\%$ before 0.36\,Myrs and declines to $\sim2\%$ by 0.96\,Myrs. Most subclusters found typically consist exclusively of some (but not all) true members or are an amalgamation of dispersed stars, stars which are in a real subcluster together and/or stars which are in different real subclusters. However some subclusters seen in 2D are true asterisms, consisting entirely of stars which come from the dispersed field. This is seen in Figure\,\ref{Fig_fcorrect} middle panel, where we show typically between $1.5-4.2\%$ of subclusters found in the 2D orientations consist entirely of members identified as part of the dispersed field in the 6D analysis. Interestingly, after 0.9\,Myrs the number of subclusters found in this category drops for D1050XY, suggesting that the additional subclusters found in this orientation at these times are less likely to be misidentified small collections of field members in the line-of-sight than at earlier stages.  
On the other hand, some real subclusters had all members assigned to different subclusters and/or the field in the 2D perspectives, i.e. they were completely missed in the 2D analysis. As shown in Figure\,\ref{Fig_fcorrect} right panel, the fraction of affected subclusters is negligible but more inclined to occur at later times, potentially due to confusion caused by the increased cluster density.

\subsubsection{Kinematics}

Figure\,\ref{Fig_vout} shows the expansion velocity as measured from our 2D/3D perspectives in the top panel, and the mean 1D expansion directional velocities in the bottom panel. Overall, the 2D perspectives generally show that the cluster is contracting. We see that the strongest contraction occurs along the $X$-axis while the cluster expands along the Z-axis for the majority of its evolution. Correspondingly the 2D perspective D1050XY (which neglects the $Z$-axis velocities) shows the strongest contraction - by 0.96\,Myrs a factor of 3.3 greater than D1050XZ which exhibits the least. As the cloud-cloud collision occurs along the $Z$-axis, these perspective differences are representative of the gas having initially higher velocities in this direction. Although the gas shocks, the velocity field is turbulent and the structure of the shock is asymmetric with non-zero velocities (to contrast with the case of uniform non-turbulent gas; see \citealt{2020MNRAS.496L...1D}). The $X$- and $Y$- axes are in the plane of the shock so their initial gas velocities are lower, hence the cluster more readily contracts in these directions and D1050XY shows the greatest contraction. Due to the asymmetry of the turbulent velocity fields, velocities in the $Y$-axis are lower than those of the $X$-axis (Figure\,\ref{Fig_vout}) which, combined with the expansion in the Z-axis, results in D1050XZ exhibiting the least contraction - while the developing shock causes a perceived expansion in this perspective until $0.63$\,Myrs.   

%#############################################
\begin{figure*}
\centering
   \includegraphics[width=0.49\textwidth]{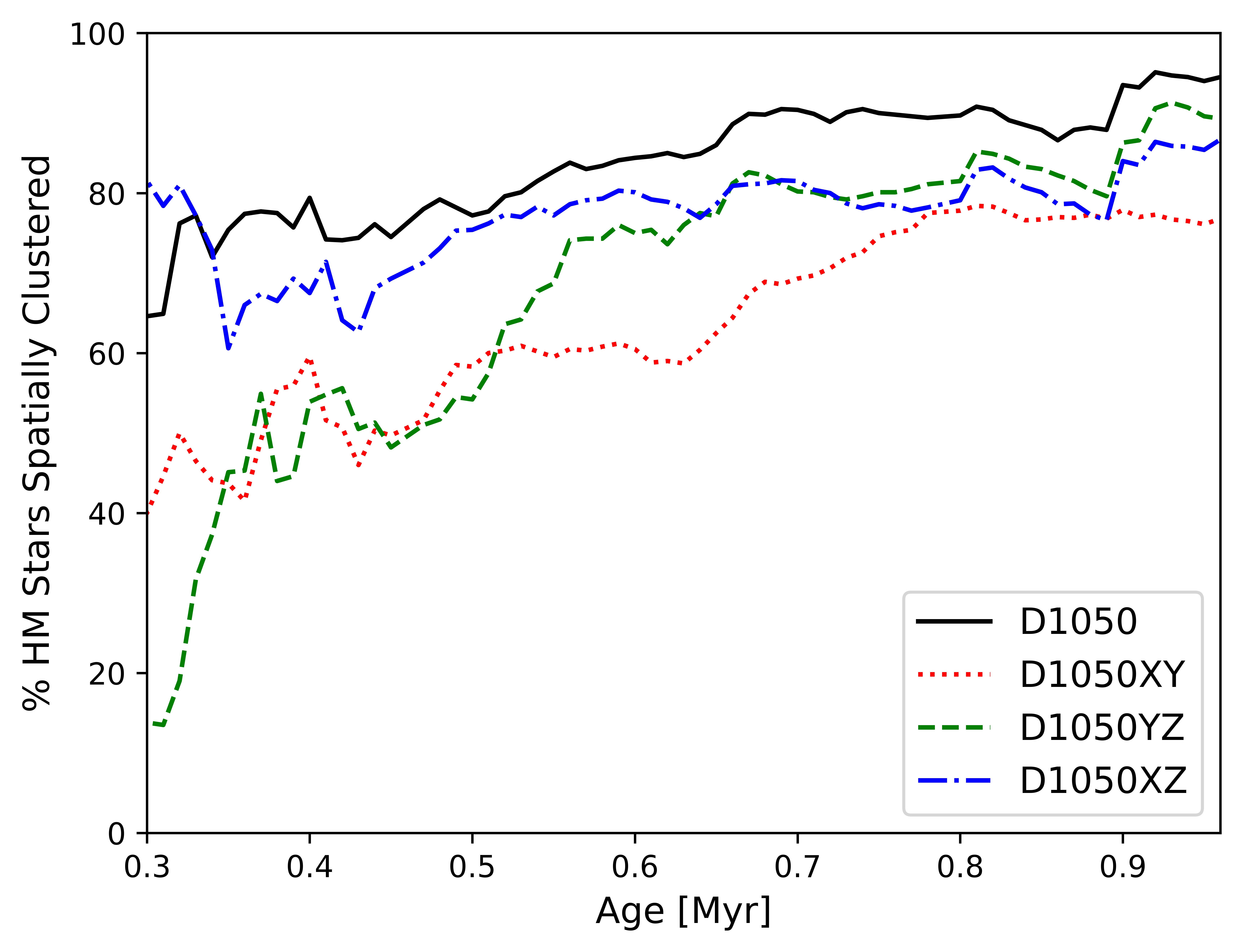}\hfill
   \includegraphics[width=0.49\textwidth]{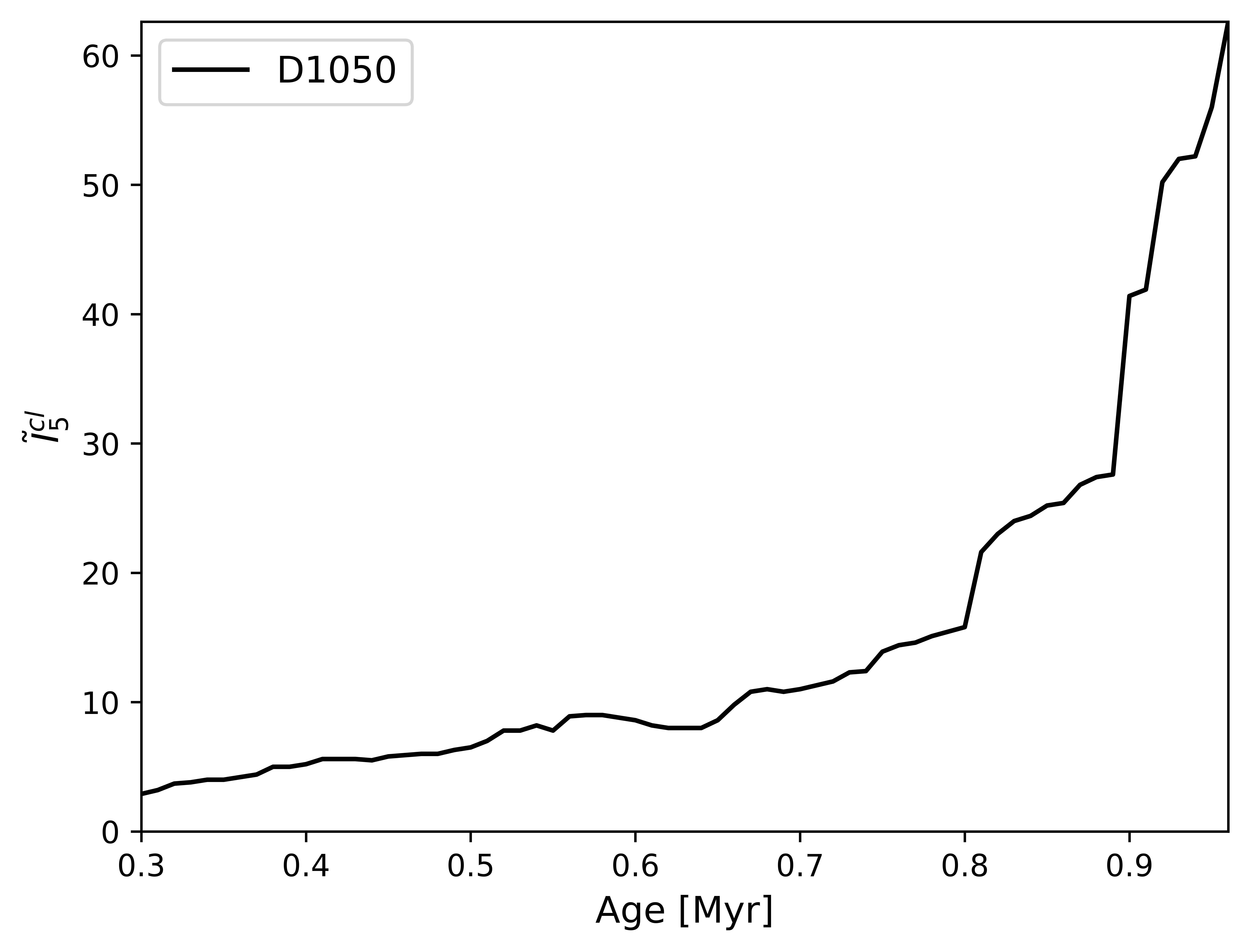}\hfill
   \includegraphics[width=0.49\textwidth]{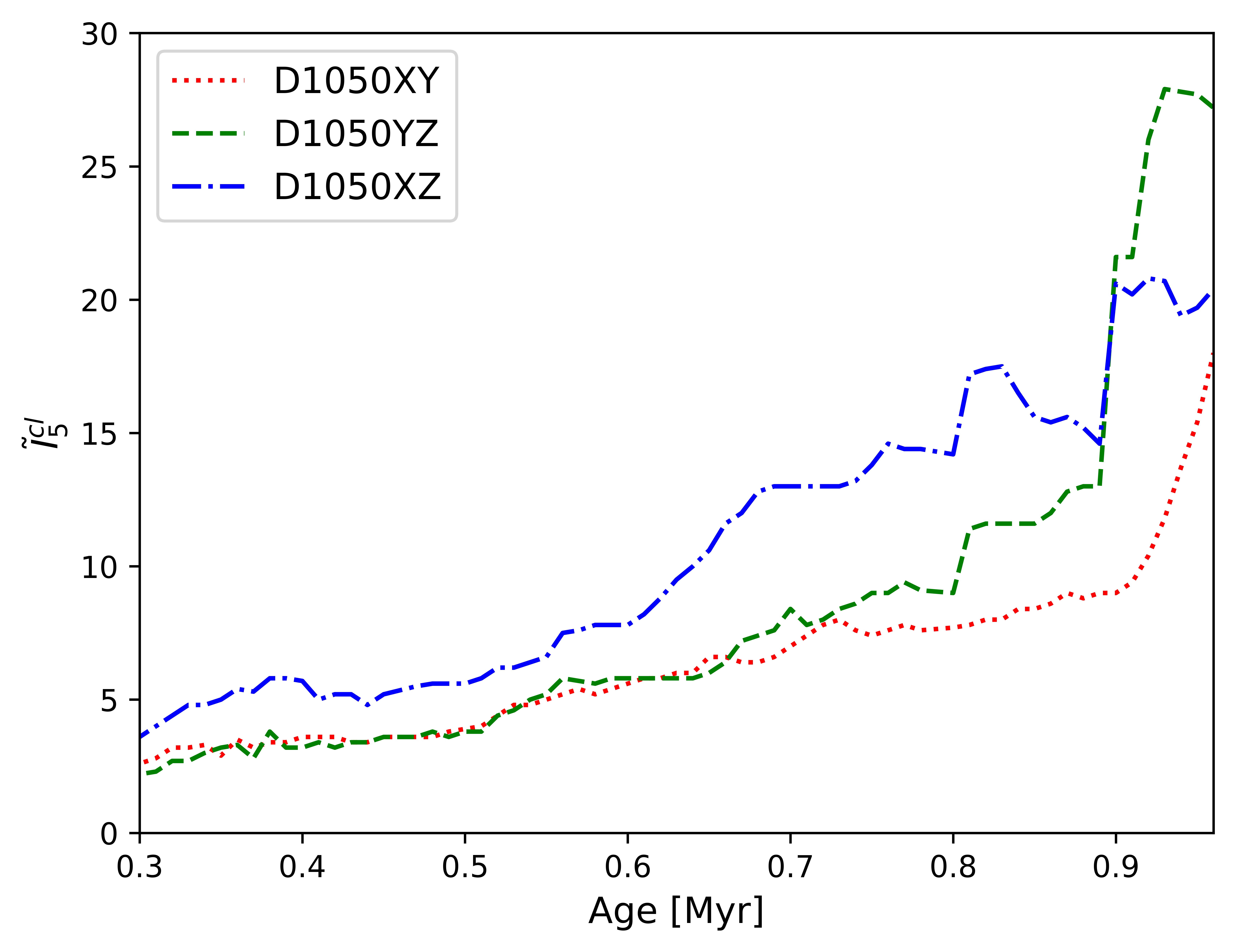}
   \caption{These panels shows signatures of Type I mass segregation found by INDICATE as a function of cluster age in 2D and 3D. (Left:) Percentage of high mass stars which are mass segregated. The strength of their spatial association is denoted by the median index value of the mass segregated stars, derived for the (Middle:) 3D and (Right:) 2D perspectives. Both the degree and strength of the segregation is underestimated in 2D  w.r.t 3D, but trends are preserved.}  \label{Fig_MS_i} 
\end{figure*}

\begin{figure*}
\centering
   \includegraphics[width=0.4\textwidth]{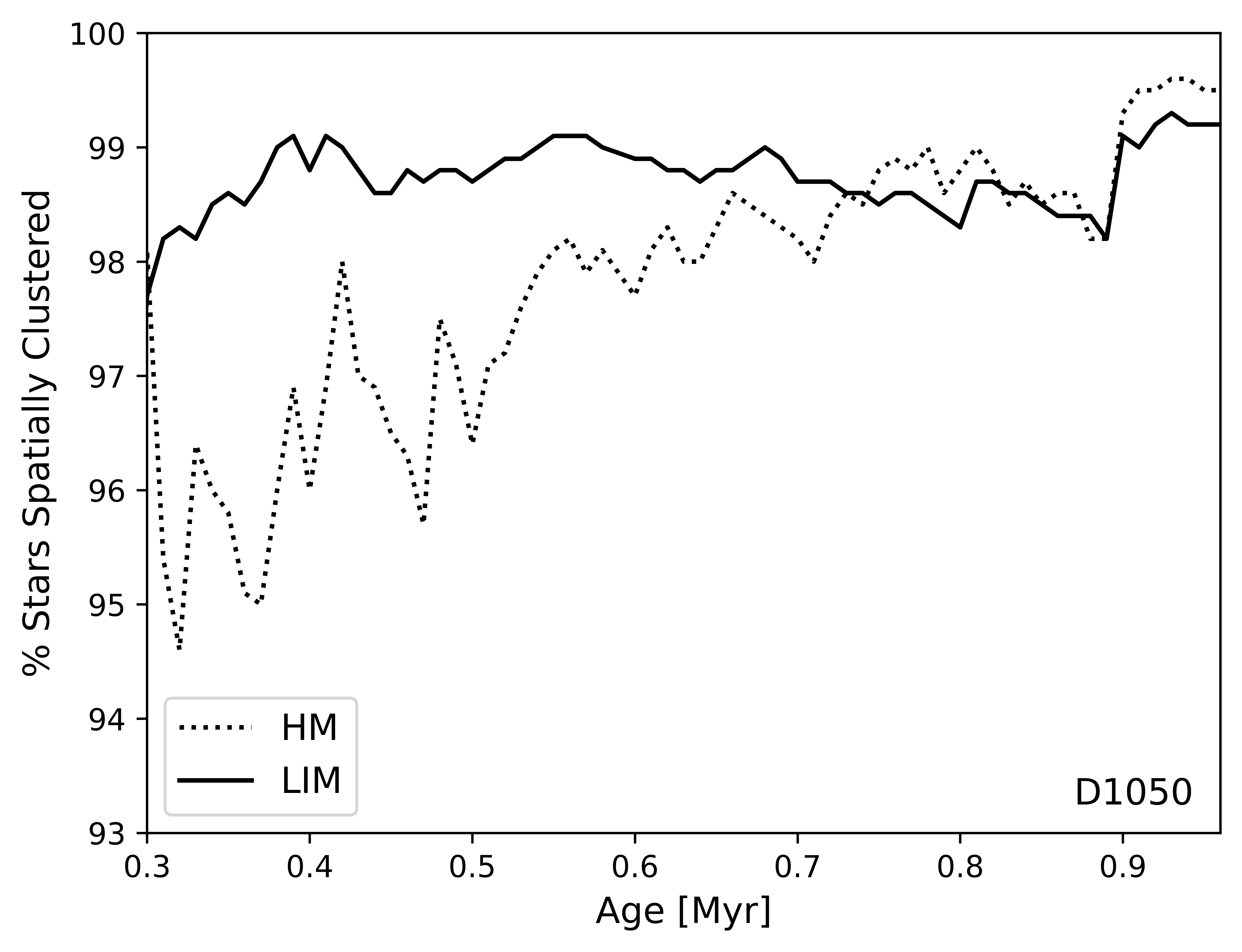}
   \includegraphics[width=0.4\textwidth]{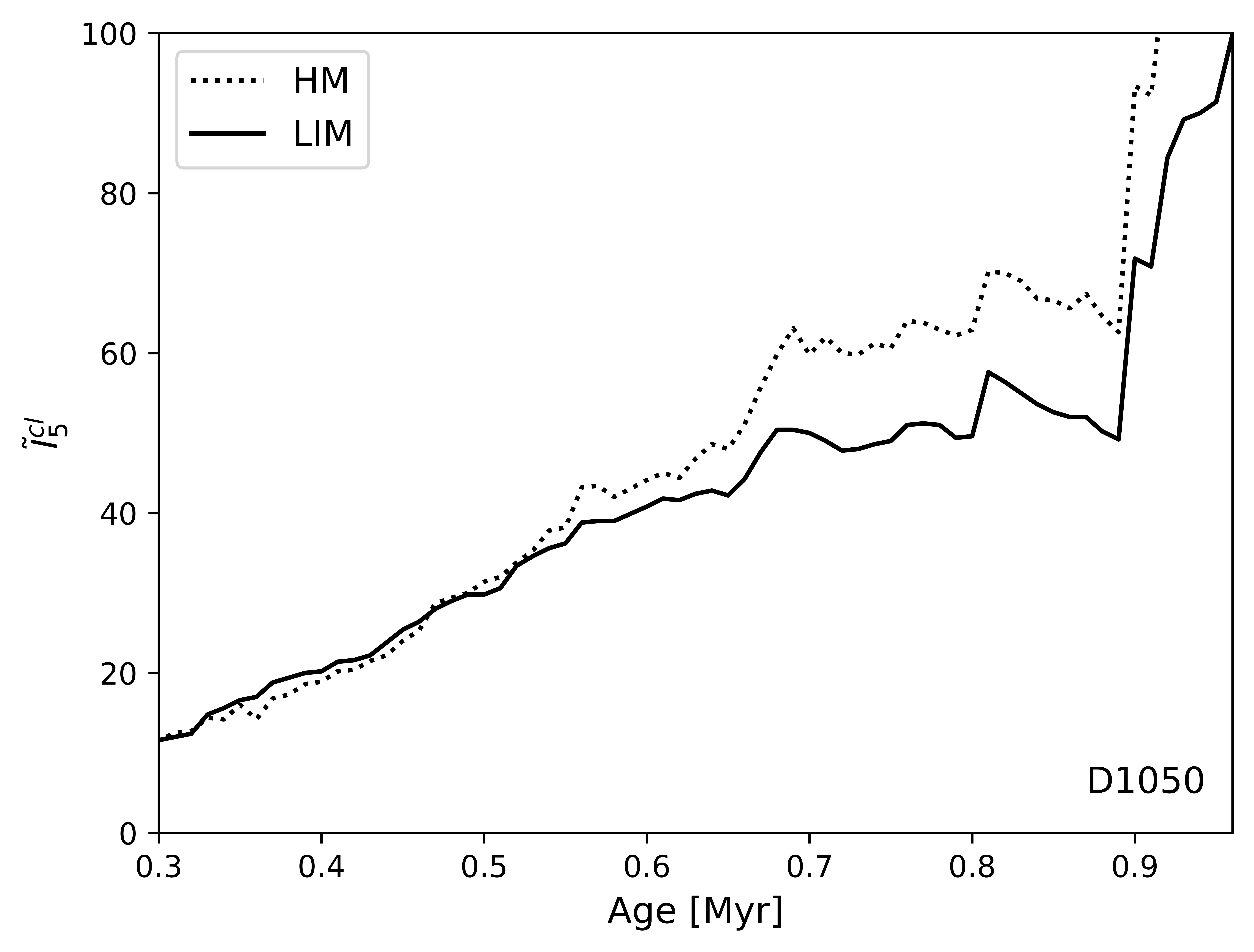}\hfill
   
   \includegraphics[width=0.4\textwidth]{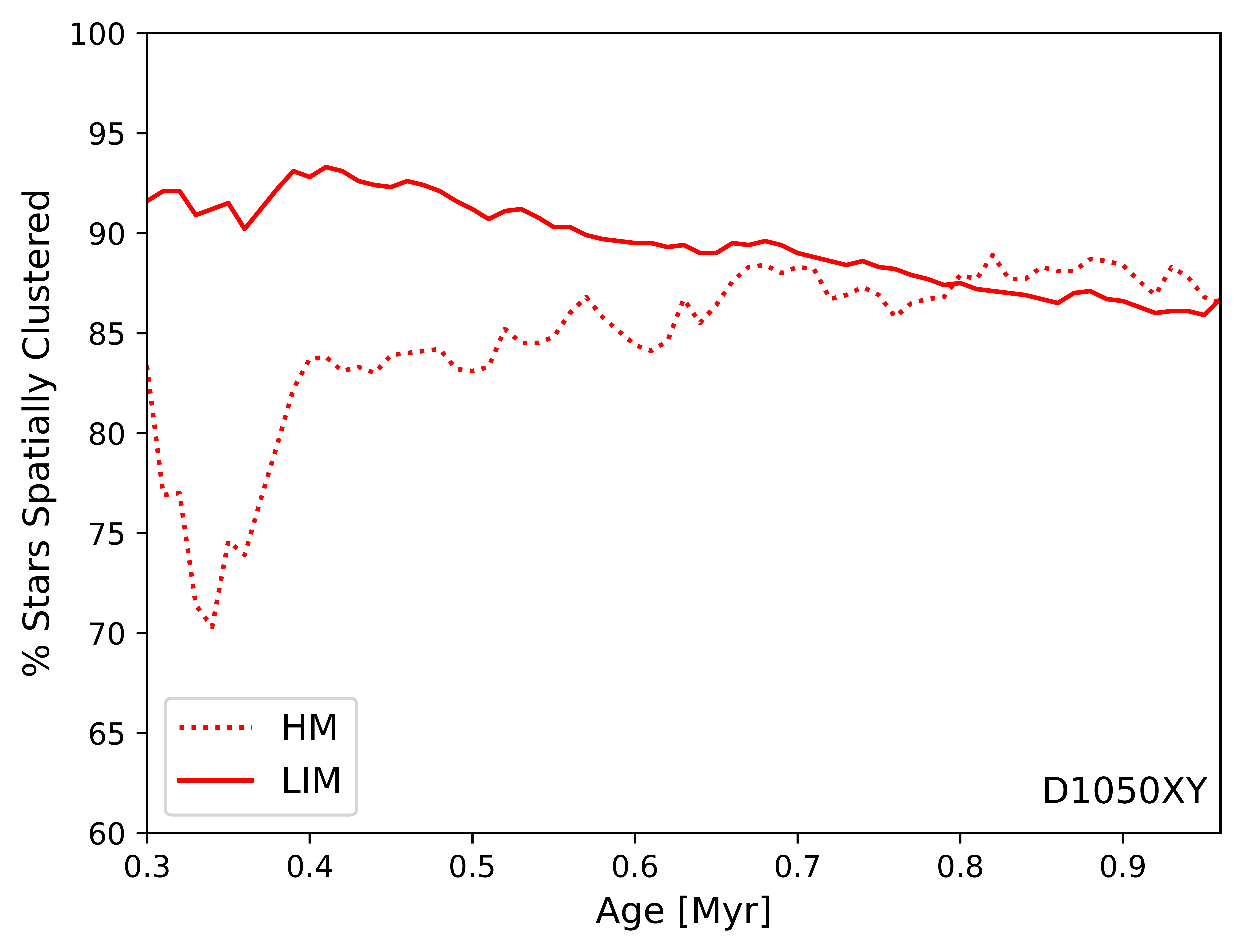}
   \includegraphics[width=0.4\textwidth]{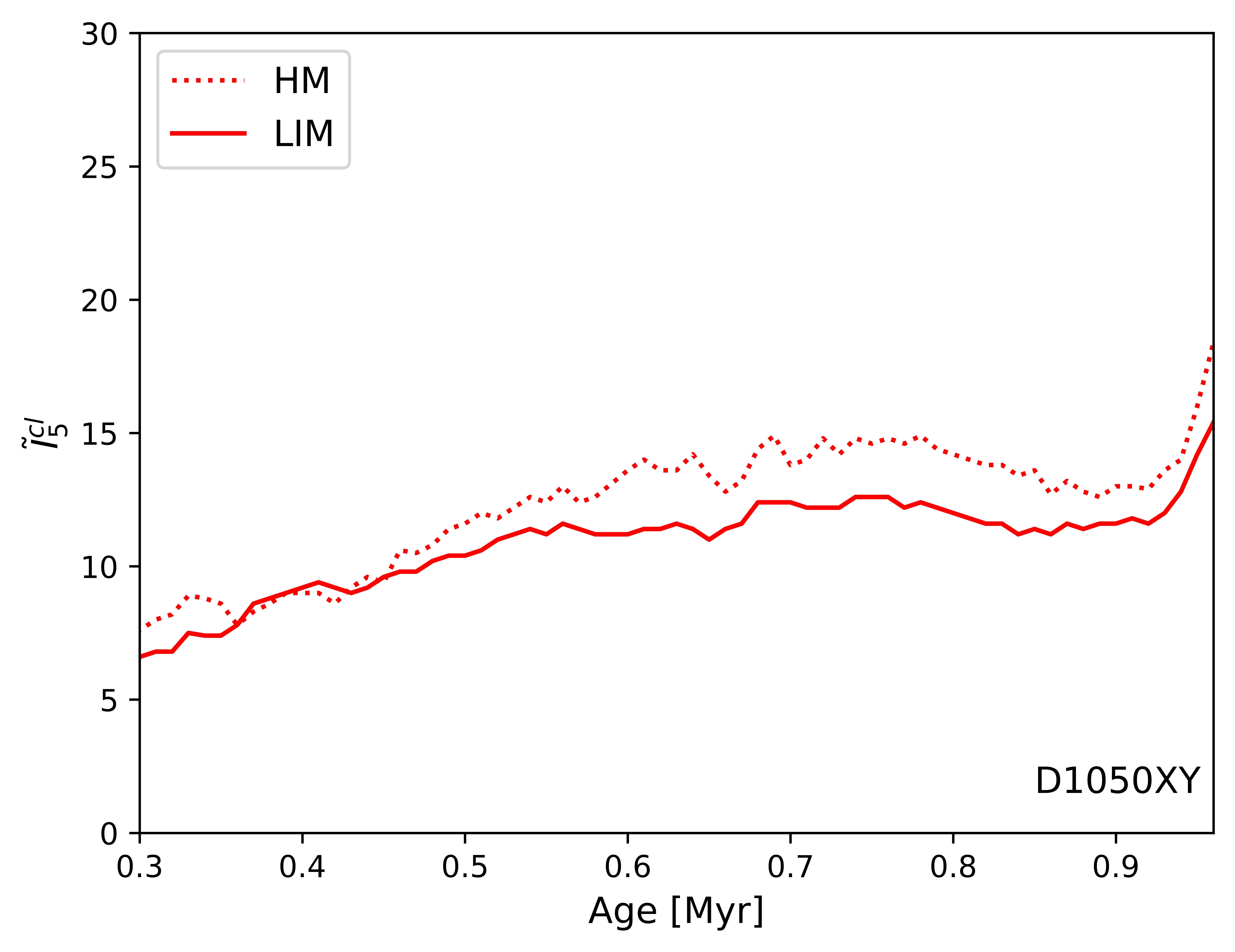}\hfill

   \includegraphics[width=0.4\textwidth]{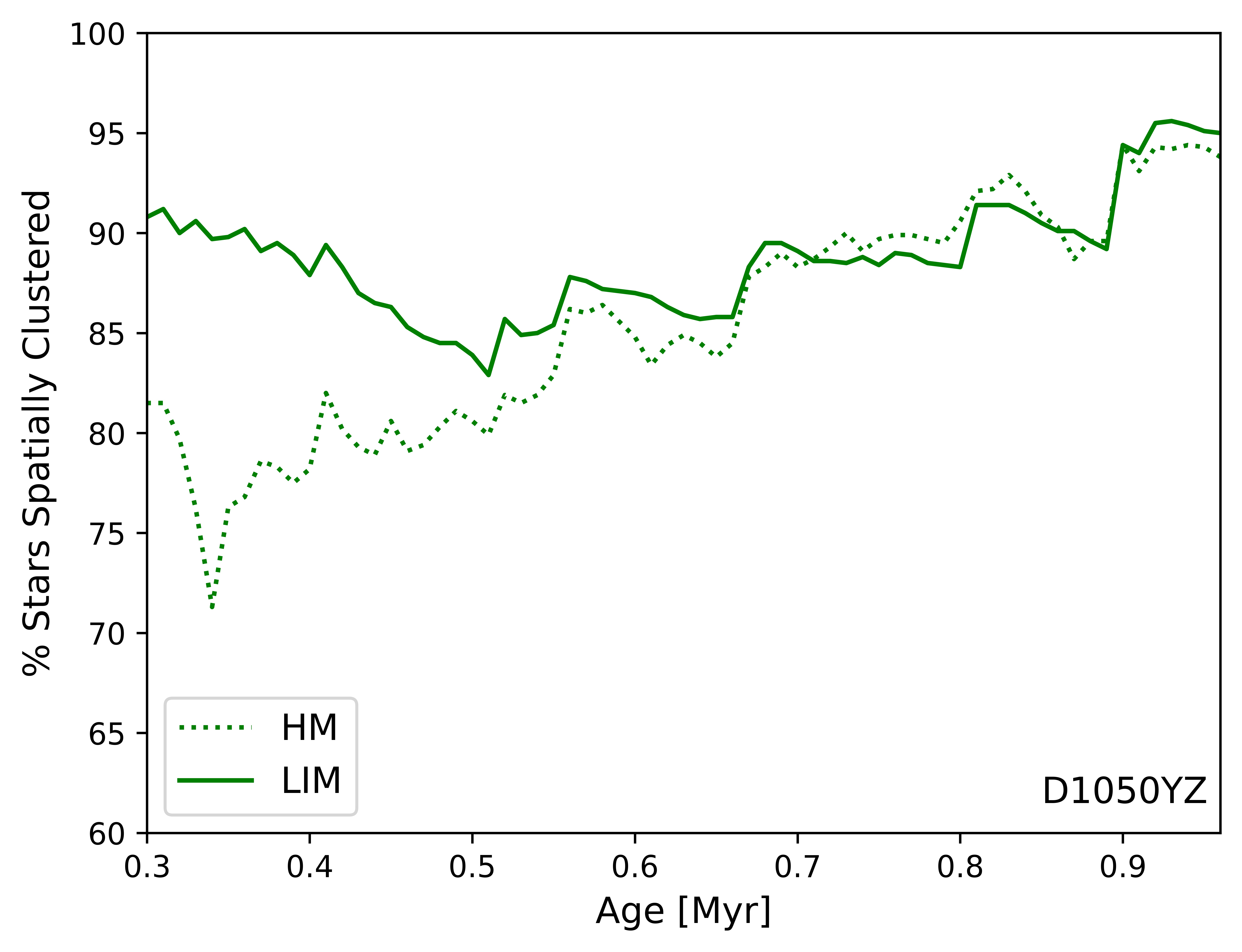}
    \includegraphics[width=0.4\textwidth]{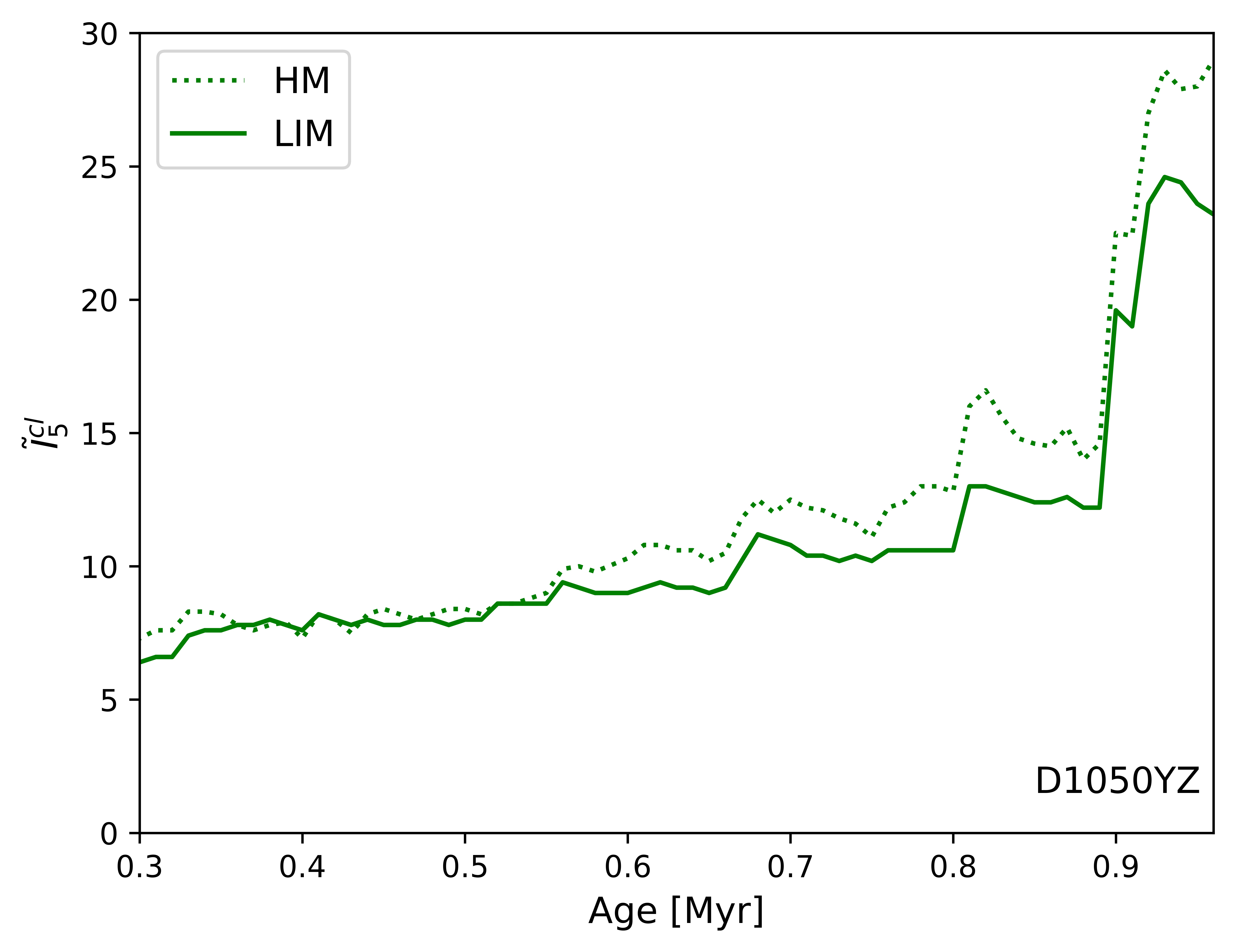}\hfill

   \includegraphics[width=0.4\textwidth]{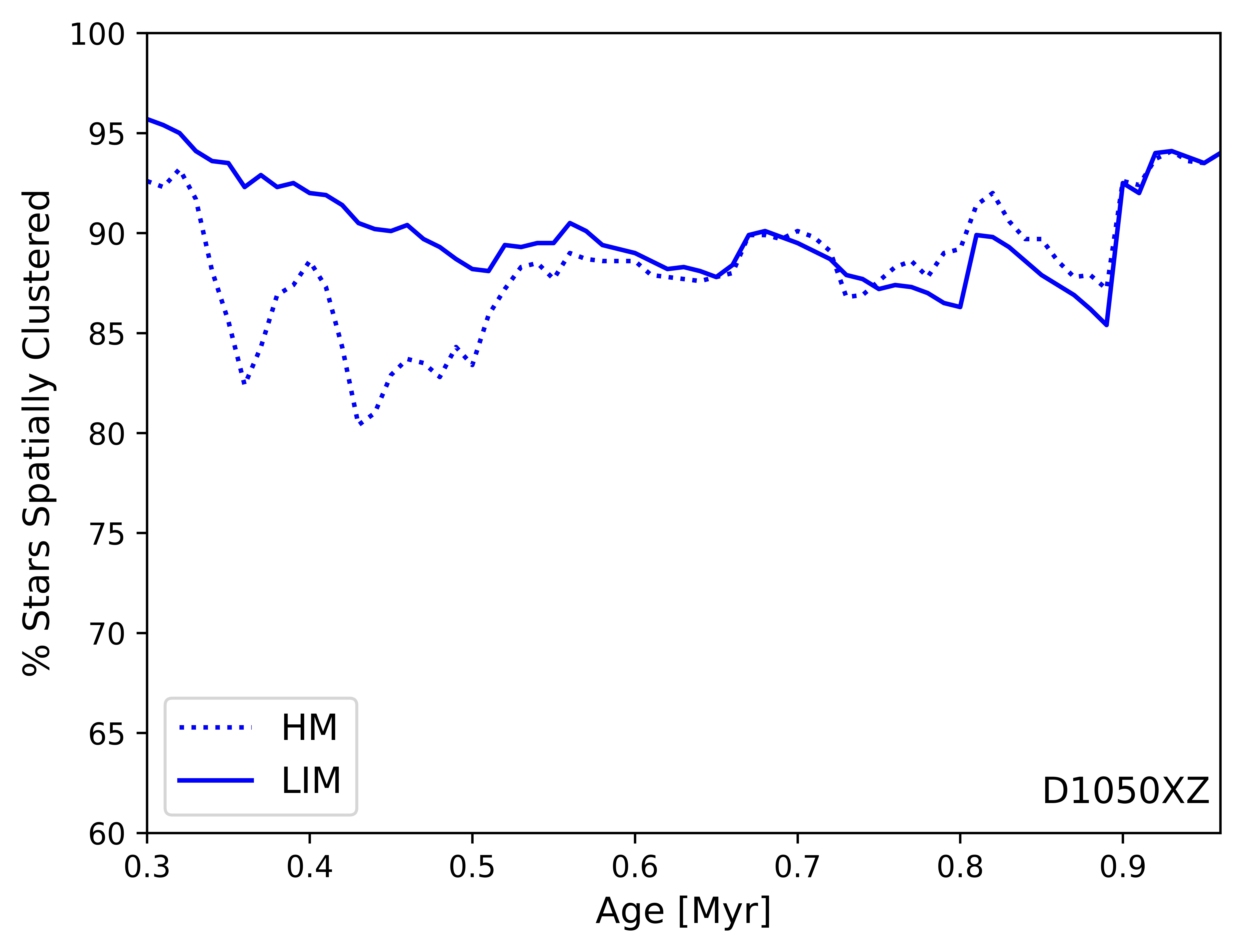}
   \includegraphics[width=0.4\textwidth]{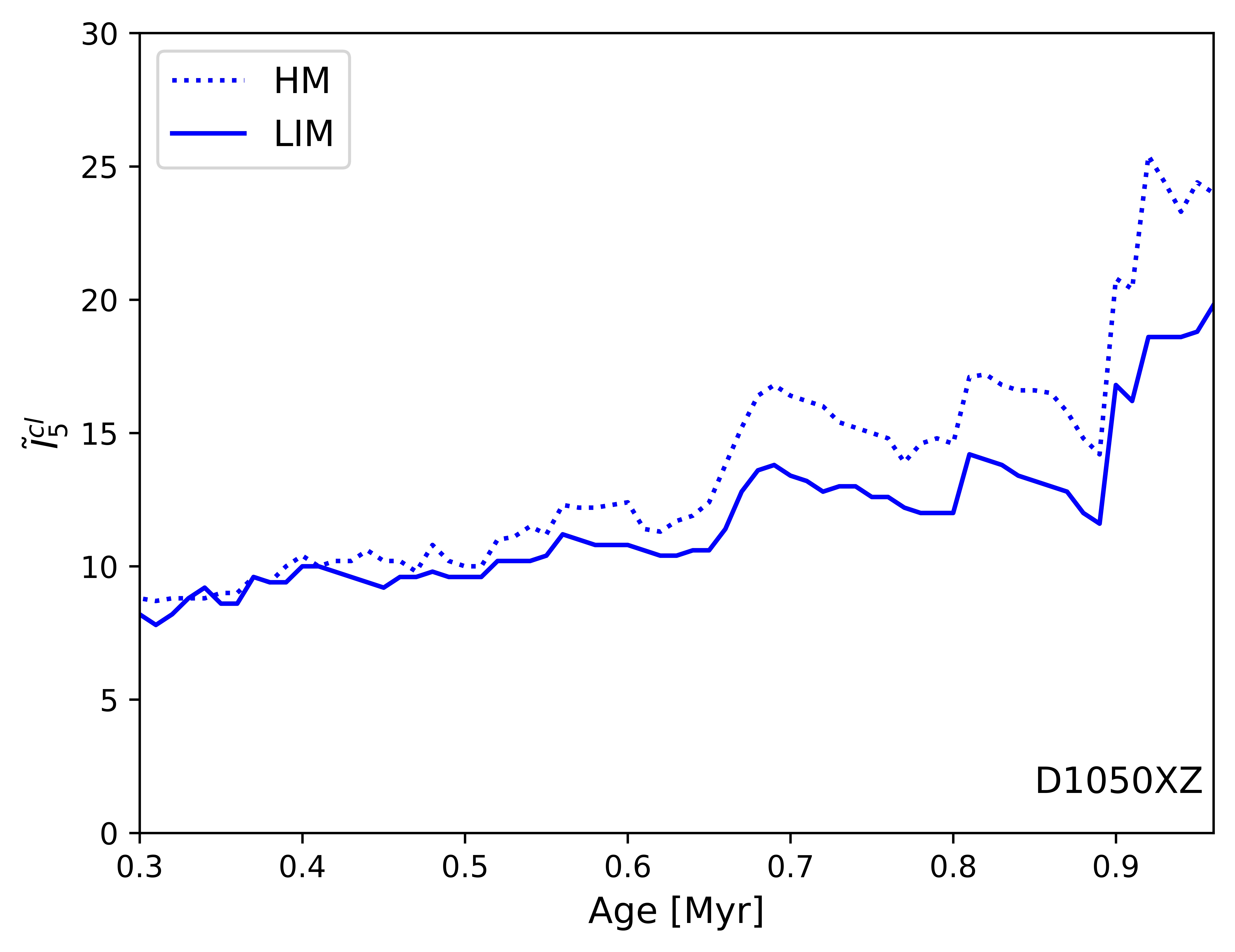}\hfill
   
   \caption{These panels show the results of the INDICATE analysis for the High Mass (HM, dotted line) and Low/Intermediate Mass (LIM, solid line) populations as a function of cluster age in (Top Row:) 3D and 2D (Upper Middle Row:) X-Y, (Lower Middle Row:) Y-Z, (Bottom Row:) X-Z planes. Plots of (Left Column:) percentage of populations identified as spatially clustered and (Right Column:) median index value of clustered stars.}  \label{Fig_MS_ii} 
\end{figure*}

%#############################################

\subsubsection{Mass Segregation}

Signatures of Type I mass segregation are weakened in all 2D orientations. Figure\,\ref{Fig_MS_i} shows that while general trends are preserved, values are consistently underestimated. The disparity between the actual and perceived fraction of segregated stars is most profound in the earlier stages of the cluster’s evolution, with the greatest disparity occurring at ages $<0.4$\,Myrs. This issue is particularly apt for D1050XY and D1050YZ where initially as few as a quarter of clustered stars ($<50\%$ total population) are identified as mass segregated compared to the 3D case, most likely due to the highly clumpy nature of the stellar distribution at this stage and its prominence in these orientations. Consequently the cluster would be incorrectly interpreted as initially having no/very weak signatures of mass segregation at $<0.4$\,Myrs when viewed from these perspectives. Furthermore, we find that the degree of association between high mass stars is consistently determined to be about half that of the 3D analysis for all 2D perspectives and time-steps. Examination of the 2D positions of massive stars correctly identifies that they are forming in relative isolation away towards the outer edges of the high mass spatial distribution, but not that they are ultimately clustering into an asymmetric clumpy amalgamation ($Q\approx0.8-0.9$ for $T>0.92\,$Myrs in D1050YZ and D1050XZ, $T>0.84\,$Myrs in D1050XY). However, as trends in the degree of association are consistent with the 3D perspective, and the majority of the population are perceived to be clustered, at ages $>0.4$\,Myrs the cluster would be correctly interpreted to possess signatures of mass segregation.

 Figure\,\ref{Fig_MS_ii} shows the general trends of the 3D Type II mass segregation analysis are preserved in the 2D perspectives but signatures are weakened. In all orientations the majority of massive stars are in stellar concentrations and initially this is proportionally less than for the LIM population, becoming equivalent between $\sim$0.65-0.80\,Myrs depending on orientation. The proportion of the two populations which are clustered at any given time is consistently underestimated in 2D (by how much varies with orientation). Evolutionary features of the tightness of this clustering for the two populations are preserved as initially they have similar degrees of association, then between 0.4-0.6\,Myrs (depending on orientation) massive stars gain relatively more intense spatial associations, and a rapid increase for both populations occurs after 0.9\,Myrs. Again, while specific values are mostly consistent between the different orientations there is some variation, and the values can be up to 8.5 times smaller than those obtained in the 3D analysis. A 2sKST confirms that once equivalent proportions of the two populations are clustered is achieved the perceived degree of association for the massive stars is significantly greater than the LIM stars in all three orientations. Therefore we find that broadly the correct conclusions are reached in the 2D perspectives: initially no mass segregation is present and high mass stars are more likely to form in isolated areas than their lower mass counterparts but do typically form in areas of comparable stellar concentrations. Later, most massive stars are mass segregated but there is a minority that are part of the dispersed population. The small variations in the age at which changes in the spatial behaviour occur are attributable to differences in the perceived stellar distribution at different cluster orientations.

\begin{figure*}
\centering
   \includegraphics[width=0.49\textwidth]{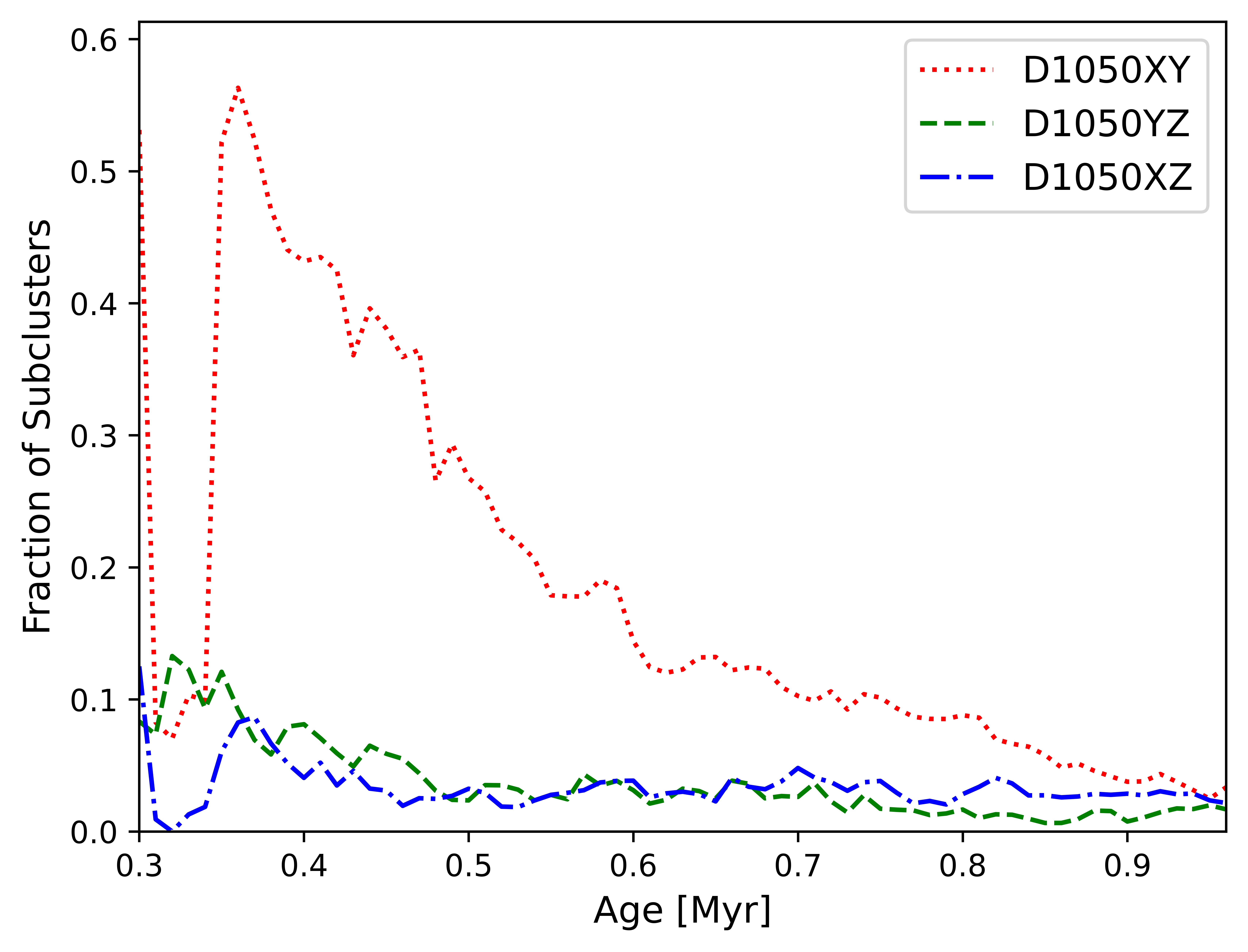}\hfill
    \includegraphics[width=0.49\textwidth]{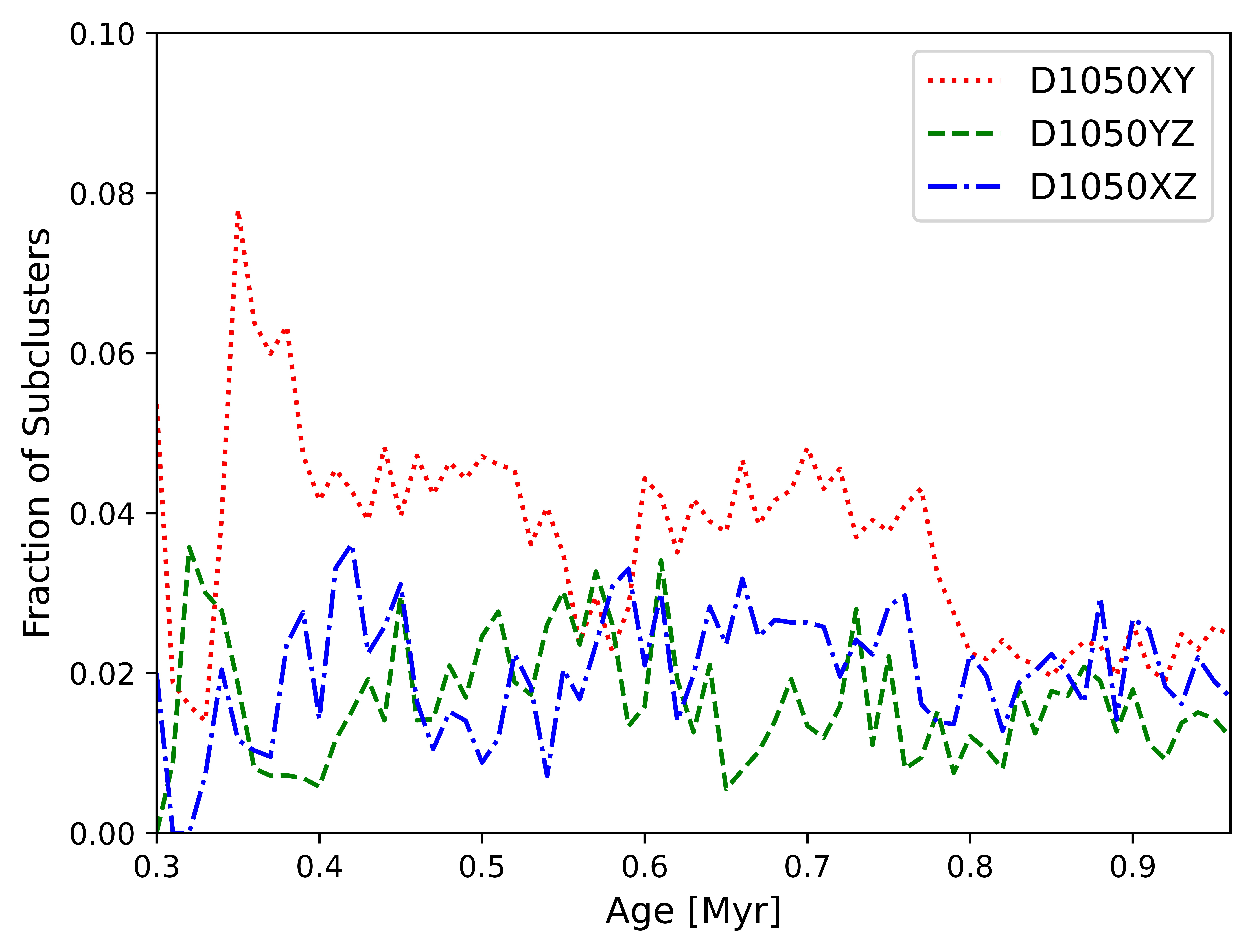}\hfill
    \includegraphics[width=0.49\textwidth]{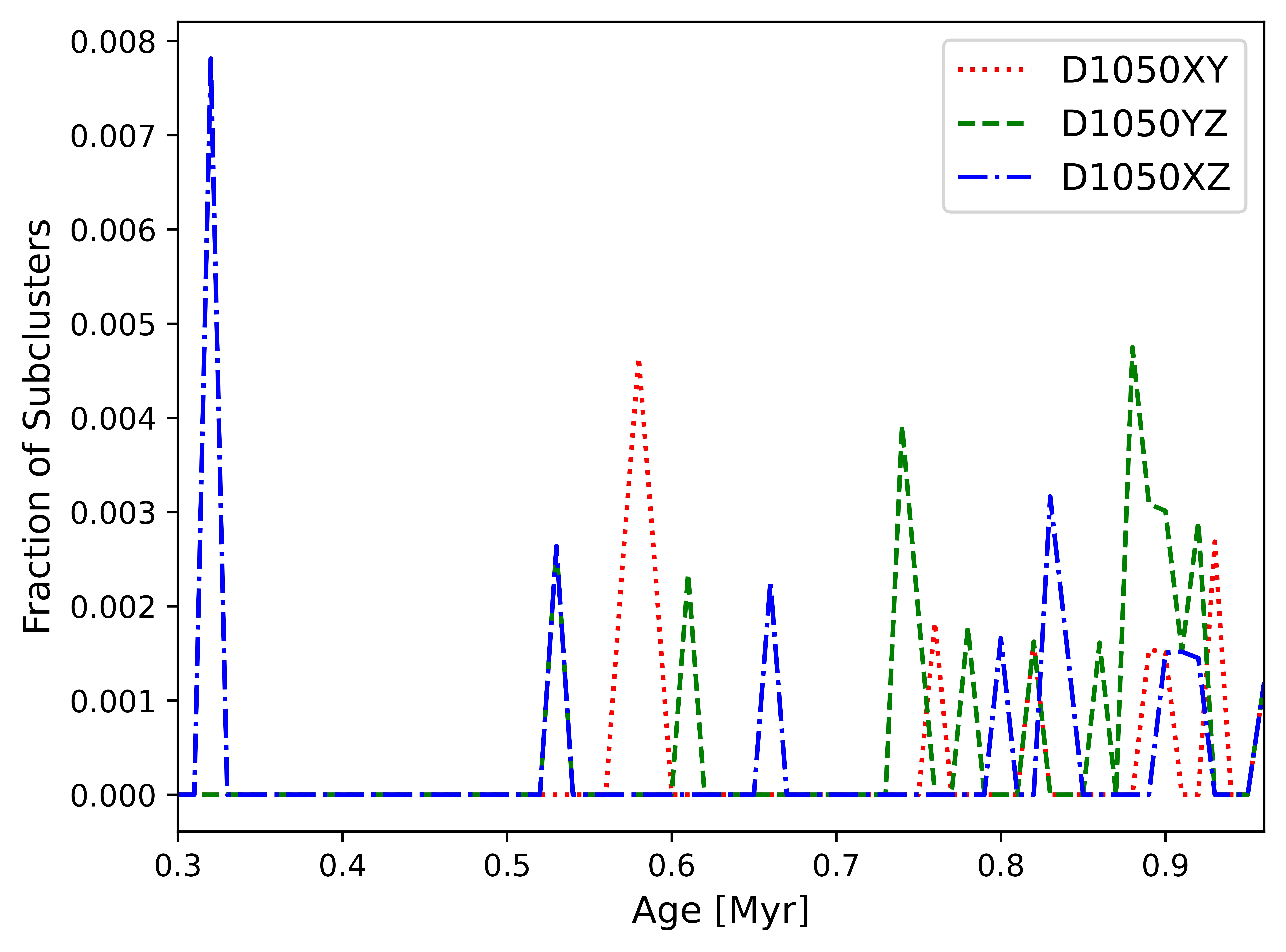}
   \caption{Plot of fraction of subclusters (Left:)  found in 2D that were also found in 3D with the same members; (Middle:) found in 2D for which all members were identified as part of dispersed population in 3D; (Right:) not found even partially in 2D that were present in the 3D data.}  \label{Fig_fcorrect} 
\end{figure*}
%#############################################

\subsubsection{Spatial Structure}

In all 2D orientations, the apparent density gradient of the cluster is correctly identified as undergoing a steady increase as the cluster evolves but remaining in a fractal distribution ($\mathcal{Q}<0.8$). However, the degree of fractal substructure appears reduced.

The lower panel of Figure\,\ref{Fig_oring_comb} shows the 2D O-ring functions.
The shortest length-scale shown (0.01\,pc) corresponds to the size of a sink.
At the longest scales the curves end when the number of particles becomes too small ($<$50) to yield meaningful results.
A good illustrative curve is the X-Y projection of D1050 at 0.3\,Myr shown as the lower blue curve in bottom panel of Figure\,\ref{Fig_oring_comb}.
The gradient changes between $10^{-2}$ and $10^{-1}$\,pc, and again around 1\,pc.
The latter represents the edge of the whole simulation, whilst the former between $10^{-2}$ and a few times $10^{-2}$\,pc is the spatial scale of the sub-clusters.
The reason for the flat region on spatial scales around 0.3\,pc and rather sharp gradient change at $10^{-1}$\,pc, both of which are largely absent from the 3D results, is the projection into 2D of many stars unrelated to the sub-cluster being measured.
We can see this projection effect is much more serious in the X-Z plane where the overall surface density is higher; it raises the surface density of unrelated stars to be comparable to that of the sub-clusters on which they are superimposed.
This effect also becomes more significant as the simulation evolves to higher density, and the density contrast becomes weaker.
\cite{bate_msdc} made a similar point, that one should expect a gradient change in the O-ring function when the density of objects of interest falls to the density of background objects.
Despite this there is still a change of gradient between $10^{-2}$ and a few times $10^{-2}$\,pc in 2D, which corresponds to the spatial scale seen in 3D, showing that even in these rather adverse circumstances the O-ring function is correctly locating the sub-structure scale.

%~~~~~~~~~~~~~~~~~~~~~~~~~~~~~~~~~~~~~~~~~~~~~~~~~~~~~~
\section{Discussion and Conclusions}
 With the recent early release of the third instalment of the Gaia survey (EDR3; \citealt{2021A&A...649A...1G}), increasing numbers of clusters have full 3D kinematic information.
 However there are still many young clusters for whom the high extinction from their natal clouds is impenetrable for optical surveys, meaning infrared data (which often lack the 3D kinematics for members) must be used instead.
 Three dimensional positional information is even rarer, since Gaia parallaxes usually cannot place a star radially within a cluster to the precision required.
 As such 2D spatial-kinematic analysis remains a popular method, and sometimes is the only viable option, to determine cluster properties. It is therefore imperative to gain a better understanding of the impact of perspective effects and how representative the properties derived in these studies are of the true cluster properties. In this work we have explored this issue by deriving the properties of a simulated cluster from the 6D data and comparing with those derived using only (2+2)D data for the cluster in different orientations. 
 
 The cluster we study here is formed via a cloud-cloud collision. We analyse this cluster partly because the collision naturally induces asymmetry in the resulting cluster making it a clear test of differences between 2D and 3D. We also check our results with a second example simulated cluster. Our analysis shows that the same qualitative conclusions tend be reached when viewing clusters in 2D versus 3D, but that quantitative conclusions when viewing in 2D are likely to be inaccurate.
 
 Arguably the greatest divergence between the true properties of the cluster and those derived using 2D data occurs in the perceived kinematics. As a consequence of the cloud- cloud collision, from which the stars are formed, the cluster is collapsing along the X- and Y- directions and expanding in the Z-direction for the majority of its traced evolution. Subsequently as the LoS orientation of the cluster is altered, contradictory determinations as to whether the cluster is expanding or contracting are made. In reality, the cluster is undergoing overall collapse between 0.3-0.96\,Myrs, which is correctly identified in the XY and YZ planes (albeit with variation in the collapse velocity), but in the XZ plane the cluster is perceived to expand until $\sim$ 0.65\,Myrs. This suggests that in the absence of 3D velocities, caution is warranted when evaluating cluster expansion/contraction from proper motions alone. This is particularly apt for young clusters which haven’t yet undergone dynamical relaxation (have an uncertain 3D shape) and/or their initial formation mechanism is unknown (differential expansion/contraction along each axis may be present).
 
 Prior to $\sim$ 0.9\,Myrs the dominant mode of star formation in our cluster is through the  collision, and while the stellar density steadily increases as the cluster evolves it remains relatively low. During this phase, despite striking differences in the visual appearance of the cluster in each orientation, the apparent spatial association of stars, changes in the degree of subclustering, perceived stellar distribution (fractal, random or radial density gradient) and mass segregation present are consistent across 2D perspectives and 3D trends are also preserved (albeit with alteration to/variation in specific values in both cases). The main discrepancies in properties determined during this phase in 2D from those of 3D, are: (i) the apparent fraction of the various populations (general, massive and low/intermediate stars) spatially clustered and how this fraction evolves with the cluster; (ii) the number of perceived subclusters, which can also dramatically differ depending on 2D orientation perspective and (iii) stellar memberships of subclusters. The majority of subclusters found in the 2D data are wrongly inferred (asterisms, amalgamations of real subclusters and/or dispersed population members, individual real subclusters identified as multiple subclusters, only partial membership of real subclusters found etc.). While a negligible fraction of real subclusters are completely missed, the fraction of real subclusters correctly identified decreases as the cluster evolves to around $4\%$ by 0.9\,Myrs. 
 
 Post 0.9\,Myrs, gravitationally driven star formation becomes significant, which together with collision-induced star formation increases the overall star formation rate and the stellar density rapidly increases. During this phase, the relative accuracy and consistency of properties derived from different orientations is reduced. 
 For example the subcluster spatial scale becomes difficult to identify against the increasing background density.
 While trends in the apparent spatial association and perceived distribution of stars are generally preserved across the 2D perspectives, and there is less discrepancy between the number of perceived subclusters in the 2D orientations, the accuracy of subcluster stellar membership continues to decline. The latter can most easily be demonstrated from \citet{Balfour2015} figures 2 and 3, who perform simulations similar to ours but with lower turbulence. From one perspective the gas forms a network of filaments, but from the edge on perspective the shock interface spans the dimension of the clouds (similar to X-Y versus X-Z in Figure~\ref{Fig_simulation}).
 
 We chose to limit our 2D analysis to just three perspectives: the XY, YZ and XZ planes, despite that clusters can potentially be in any orientation in the LoS of an observer. Orientating the cluster to have a LoS in-between the planes therefore gives it an apparent shape which is a ‘hybrid’ of these three extremes. We also chose a relatively extreme case; differences from 2D perspectives may likely be smaller generally. Additional observational biases, such as sample incompleteness, were not considered in our analysis as these are outside the scope of this work. The purpose of the present study was to assess the impact of 2D projection on derived cluster properties, so by not introducing additional biases our results could be definitively attributed to 2D projection effects, rather than any specific set of observing conditions and/or combination of biases.  These issues will be the focus of upcoming papers in this series. Overall we find that local environmental conditions of a cluster play a role in the accuracy of derived cluster properties so in the absence of 6D data these should be routinely taken into consideration when undertaking, and care should be taken when interpreting and assigning significance to the results of, 2D cluster analyses.

%~~~~~~~~~~~~~~~~~~~~~~~~~~~~~~~~~~~~~~~~~~~~~~~~~~~~~~
\section*{Data Availability}
The simulation data underlying this article was provided by KYL by permission, which will shared on request to the corresponding author with permission of KYL. The analysis data underlying this article will be shared on reasonable request to the corresponding author.

%~~~~~~~~~~~~~~~~~~~~~~~~~~~~~~~~~~~~~~~~~~~~~~~~~~~~~~
\section*{Acknowledgements}
AB, CLD and TN are funded by the European Research Council H2020-EU.1.1 ICYBOB project (Grant No. 818940). SR acknowledges funding from STFC Consolidated Grant ST/R000395/1 and the European Research Council Horizon 2020 research and innovation programme (Grant No. 833925, project STAREX). The authors would like to thank the referee R. Parker for his constructive and insightful feedback which led to the improvement of the manuscript.

%~~~~~~~~~~~~~~~~~~~~~~~~~~~~~~~~~~~~~~~~~~~~~~~~~~~~~~
%#############################################

\bibliographystyle{mnras}

\bibliography{Buckner_OBYMCC1.bib}
\bsp

%#############################################
\newpage
\appendix
\section{Summary for C1020}\label{results_c1020}

To check whether our findings for the effects of 2D perspective on cluster characterisation are not unique to our chosen cluster, we repeat our analysis on a second simulation, which is also the `L20Turbulent' model in \cite{liow_grouped_2021}.

\begin{figure*}
    \centering
    \includegraphics[width=0.9\textwidth]{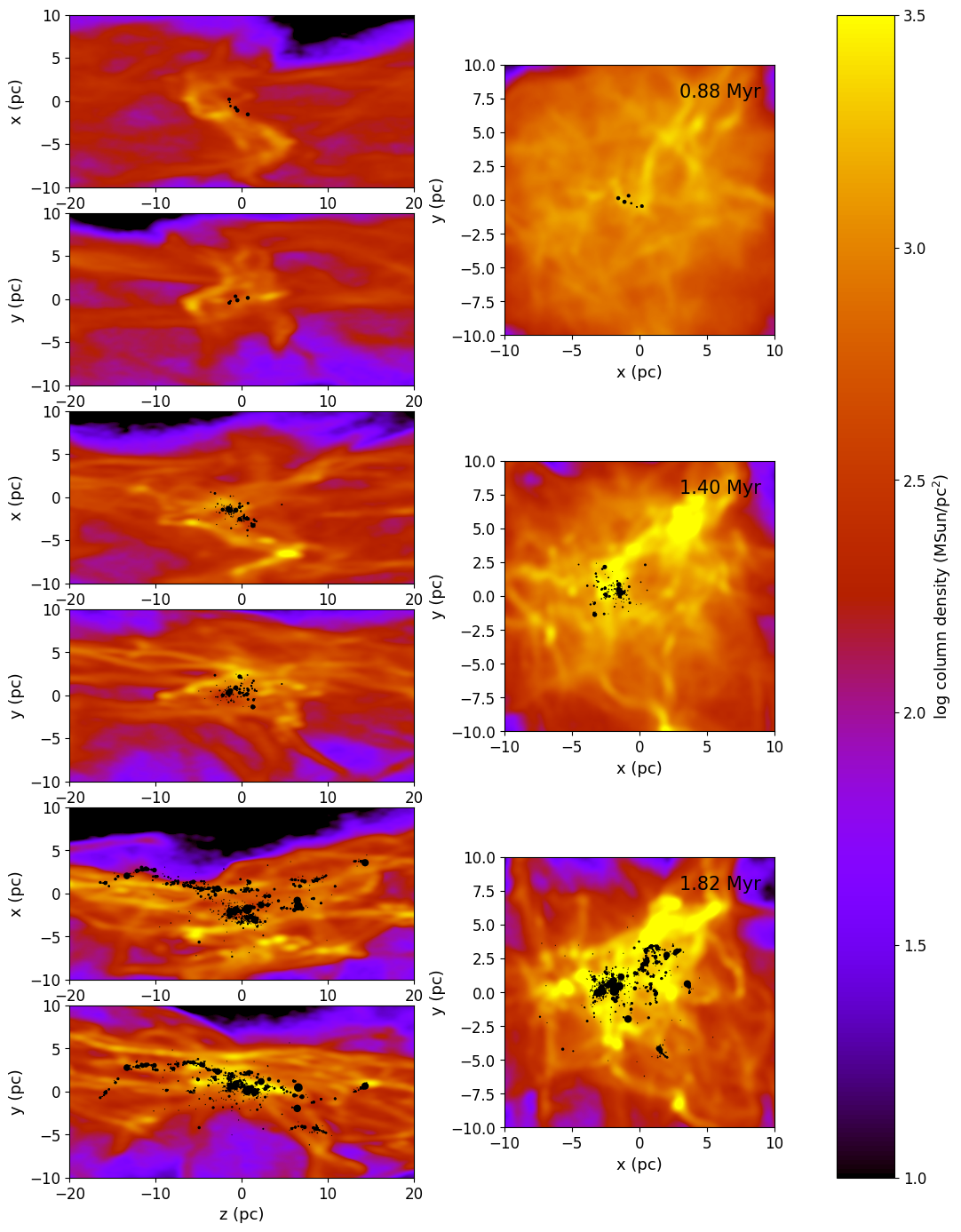}
    \caption{This figure shows our simulated YMC C1020 at 3 different stages of evolution and 2D perspectives. The cluster is
    formed from a z-axis cloud-cloud and is shown in the X-Z, Y-Z and X-Y planes at an age of (Top row:) 0.88\,Myr, (Middle row:) 1.40\,Myr and (Bottom row:) 1.82\,Myrs. Black dots represent stars and the colour scale denotes gas density.}
    \label{fig:additional_model}
\end{figure*}

This simulation is a lower speed and lower density cloud-cloud collision model as compared to the main simulation in this paper, i.e. the initial cloud density is $1.03 \times 10^{-21}$ g cm$^{-3}$ and the relative velocity is 10 km s$^{-1}$. The parameters are mostly similar to those used in the main simulation in this paper, however the details can be found in Section 2.3.2 of \cite{liow_grouped_2021}. In this model as shown in Figure \ref{fig:additional_model}, stars are formed via collision at the compressed central region. At a later time, stars are also created due to gravitational collapse of condensed sub-regions within the clouds, causing the stellar distribution to appear filamentary. Because the clouds are elongated along the $z$-axis, these filaments are approximately parallel to the $z$-axis, causing significant asymmetry in the system.

The low density and small number of stars in this simulation make it difficult to draw any significant conclusions from the O-ring function.
Otherwise find our results for C1020 are consistent with those found for D1050. 
Specifically:\\

\begin{enumerate}
    \item As the system evolves the percentage of members spatially clustered, and average degree of association of those stars, are found to be broadly similar in 2D perspectives to the results from 3D, but at lower intensities.\\

    \item the fraction of stars in subclusters in 2D is generally consistent (with some exceptions) with that found in 3D. \\
    
    \item the number of subclusters  identified  in the 2D perspectives are increasingly overestimated with time as the system evolves (and the cluster becomes more densely populated). The degree of overestimation can vary significantly with cluster orientation. \\
    
    \item stellar memberships of subclusters are inaccurate, and accuracy declines as the system evolves. Subclusters  found  in  the  2D  data are  often wrongly  inferred, but a negligible fraction (and in this case none) of the real subclusters  are completely  missed. The number of subclusters correctly identified in C1020 is typically higher than in D1050 (a minimum of $20.4\%$ accuracy is achieved at the end evolutionary point, for example) indicating the degree of accuracy for subcluster identification is linked to the local conditions of a region.\\
    
    \item the fraction of subclusters found in the 2D perspectives with members that are actually part of the dispersed population is mostly constant as the system evolves (albeit slightly higher at $\sim 8-10\%$).\\
    
    \item depending on the orientation of the cluster, 2D determination of whether the cluster is expanding or contracting can be incorrectly inferred. If correctly determined, the perceived velocity may differ from that of the 3D data.  \\
    
    \item Signatures of Type I and II mass segregation are preserved, but specific values vary. \\
    
    \item determination of whether the cluster is in fractal or radial distribution in the 2D perspectives are accurate. However, the degree of fractal substructure appears reduced.
    
\end{enumerate}

%###########################################

\label{lastpage}
\end{document}